\documentclass{ws-ijmpd}
\usepackage{graphicx}
\begin{document}

\markboth{Yang Huang, Yi-Ping Dong, Dao-Jun Liu}
{Revisiting the shadow of a black hole in the presence of a plasma}

%
\catchline{}{}{}{}{}
%

\title{Revisiting the shadow of a black hole in the presence of a plasma}

\author{Yang Huang}

\address{Center for Astrophysics and Department of Physics, Shanghai Normal University, 100 Guilin Road, Shanghai 200234, China\\
saisehuang@163.com}

\author{Yi-Ping Dong}

\address{Center for Astrophysics and Department of Physics, Shanghai Normal University, 100 Guilin Road, Shanghai 200234, China}

\author{Dao-Jun Liu\footnote{Corresponding author}}
\address{Center for Astrophysics and Department of Physics, Shanghai Normal University, 100 Guilin Road, Shanghai 200234, China\\
djliu@shnu.edu.cn}

\maketitle

\begin{history}
\received{Day Month Year}
\revised{Day Month Year}
\end{history}

\begin{abstract}
We study the photon's motion around a black hole in the presence of a plasma whose density is a function of the radius coordinate by a renewed ray-tracing algorithm and investigate the influence of the plasma on the shadow of the black hole. The presence of plasma  affects not only the size but also the shape of  black hole shadow.  Furthermore,  the influence  of plasma on trajectories of photons depends on the frequency of photon. For the high-frequency photons, the influence is negligible, on the contrary, the trajectories of low-frequency photons is affected significantly by the plasma. Interestingly, it is also found  that the black hole image would  take on a  multi-ring structure due to the presence of plasma.
\end{abstract}


\ccode{PACS numbers:}


\section{Introduction}

Black holes  are the most fascinating objects in the universe. 
In the vicinity of a black hole, due to its extremely  strong gravitational field, light rays are strongly deflected and can even orbit around the black hole for many laps  before continuing on their way. An interesting question is  what black holes look like to a distant observer. 
Direct observation of an astrophysical black hole is not possible, because of the existence of the event horizon. Moreover,  the relatively small size of black holes and their  large distance from us make them very difficult to observe. 
However, when we observe a black hole directly, it is expected to see a silhouette in the sky, which is commonly called the \textit{shadow} of the black hole.  
Theoretically, the shadow is defined as the region of the observer's sky that is left dark if there are light sources distributed everywhere but not between the observer and the black hole. They are determined by the black hole absorption cross section for light at high frequencies. Moreover, the size and shape of the shadows are sensitive to the parameters of black hole, such as mass and spin, as well as the orientation of spin axis of the black hole with respect to the observer.

It is anticipated that the shadow of some super-massive black holes, such as Sagittarius A* (Sgr A*) at the center of our galaxy, the Milky Way as well as  M87 in the center of the Virgo A galaxy may be observable with very long baseline interferometry (VLBI) at sub-millimeter wavelengths  \cite{Falcke2000,Doeleman:2008qh}, assuming that the accretion flow is optically thin in this region of the spectrum. Observations of the shadow will not only provide compelling evidence for the existence of event horizon, but also place the estimates on the black hole parameters \cite{PhysRevD.80.024042}.

Therefore, it is important to investigate analytically or numerically the shape of black hole shadow in different configurations or circumstances, thereby providing new templates for the ongoing observations. 
The shadow cast by a Schwarzschild black hole is a circle, the radius of which  was first calculated by Synge \cite{Synge1966} long time ago. The shadow of a rotating Kerr black hole, which is most relevant in astrophysics,  is no longer a circle but an elongated silhouette in the direction of the rotation axis due to the dragging effect \cite{Bardeen1973,Chandrasekhar1983,Takahashi2004}. In recent years, much attention has been paid to the apparent shape of the rotating non-Kerr black holes, such as  black holes with a NUT-charge \cite{Chakraborty2014,PhysRevD.89.124004},  with scalar hair \cite{PhysRevLett.115.211102,Cunha2016a}, with quadrupole mass moment \cite{Wang:2018eui} and those in alternative theories of gravity, see e.g. Refs.[ \refcite{doi:10.1142/S0218271809014881,Amarilla2012,Amarilla2013,Wei2013,Li2013,Moffat2015,CUNHA2017373,Vetsov2018}].  
The shadows of  exotic objects such as wormholes \cite{PhysRevD.88.124019,PhysRevD.91.124020} and naked singularities \cite{PhysRevD.92.044035} have  been also calculated and compared with those of black holes.  On the other hand, rather than investigate all possible theories of gravity and their corresponding black hole solutions, some parameterized metrics representing rotating non-Kerr black holes are also investigated by calculating  their shadows \cite{Johannsen:2010ru,Younsi2016a}.

It is believed that realistic astrophysical black holes or other compact objects interact with their environment and are surrounded by ionized matter. Therefore, it is of relevance to consider the  effect of surrounding plasma on the motion of light rays in the black hole spacetime. In the book \cite{Perlick2000}, 
Perlick discussed the influence of a spherically symmetric and time-independent plasma on the light deflection in Schwarzschild spacetime and calculated the light deflection in the equatorial plane in the Kerr spacetime. 
The potential observability and applications of such frequency-dependent plasma effects in general relativity for compact objects is considered by Rogers \cite{Rogers:2015dla}. Er and Mao investigate effects of plasma on gravitational lensing \cite{Er:2013efa}. The strong gravitational lensing on the equatorial plane of the Kerr black hole is studied by Liu \emph{et al.} \cite{Liu2016}. The influence of a plasma on the shadow of a spherically symmetric black hole is investigated by Perlick \textit{et al.} \cite{Perlick:2015vta}. 

For the Kerr black hole in vacuum, the null geodesic equations are fully separable, leading to four constants of motion \cite{Carter1968}, and the problem is remarkably simplified. However, for general non-Kerr rotating black hole or Kerr black hole surrounded by ionized matter,  Carter constant does not usually exist and Hamilton-Jacobi equations for the light rays are not separable. Therefore, it is difficult to calculate analytically. Interestingly, Perlick and Tsupko  have found that  the distribution of plasma has to  satisfy some special  condition that the existence of a generalized Carter constant is guaranteed \cite{PhysRevD.95.104003}.

In this work, we shall use a renewed ray-tracing algorithm to study numerically the photon's motion around  Kerr black holes in the presence of a plasma  with radial density  and investigate the influence of  plasma on the shadow of a Kerr black hole. This paper is organized as follows. In Section \ref{sec:2}  we  present the equations of motion for the light ray in the background of Kerr black hole with a plasma. The Ray-tracing algorithm we use is introduced in Section \ref{sec:RTA}.
Then, in Section \ref{sec:results} we present the shadows for a representative sample of
solutions and interpret the obtained patterns. Finally, we conclude with a discussion in Section \ref{sec:conclusion}. Throughout the paper, the natural units $G_N=c=1$ are used.

\section{Light rays in the background of a Kerr black hole surrounded by a plasma}
\label{sec:2}
Let us consider a rotating black hole surrounded by a non-magnetized cold  plasma. The background spacetime is described by the Kerr line element, which, in standard Boyer-Lindquist coordinates, is given by 
\begin{equation}\label{Kerr Newman metric}
\begin{aligned}
ds^2=&-\frac{\Delta}{\rho^2}\left(dt-a\sin^2\theta d\phi\right)^2+\frac{\rho^2}{\Delta}dr^2\\&+\rho^2d\theta^2+\frac{\sin^2\theta}{\rho^2}\left[\left(r^2+a^2\right)d\phi-a\;dt\right]^2,
\end{aligned}
\end{equation}
where
\begin{equation}\label{rho^2 and Delta function}
\rho^2\equiv r^2+a^2\cos^2\theta,\;\Delta\equiv r^2-2Mr+a^2.
\end{equation}
Here, $M$ and $a$ are the mass and angular momentum per unit mass of the Kerr black hole, respectively. 

The photon's motion in a non-magnetized pressure-less plasma is characterized by the Hamiltonian
\begin{equation}\label{Eq: Hamiltonian}
\mathcal{H}=\frac{1}{2}g^{\mu\nu}p_\mu p_\nu+\frac{1}{2}\omega_p^2.
\end{equation}
Here $g^{\mu\nu}$ is the contravariant metric tensor and $\omega_p$ denotes the  electron frequency of the plasma, which is given by
\begin{equation}
\omega_p^2(x)=\frac{4\pi e^2}{m} N(x),
\end{equation} 
where $e$ and $m$ represent the charge and mass of electron, respectively, and $N(x)$  is the number density of the electrons in the plasma.  In this paper, the plasma density is assumed to be a function of the radius coordinate only, i.e., $N(x)=N(r)$.
To be specific, we consider two models for the distribution of plasma. The first one is the power-law model, in which the profile of  plasma density
\begin{equation}
N(r)=N_0\left(\frac{r_0}{r}\right)^h,
\end{equation}
where $N_0$, $r_0$ and $h$ three positive constants, then
\begin{equation}\label{Eq: plasma md}
\omega_p^2=\frac{k_0}{r^h}
\end{equation}
where we have let $r_0=M$ and $k_0=4\pi e^2N_0/m$.
In the second model, the profile of  plasma density takes an exponential form
\begin{equation}
N(r)=N_0e^{-r/r_0},
\end{equation}
then $\omega_p^2=k_0e^{-r/r_0}$, where $k_0$ has the same definition as in the power-law model, but $r_0$  remains unassigned.

The photon's motion in a plasma is governed by the Euler-Lagrange equation
\begin{equation}\label{Eq: EL eqs}
\frac{\partial\mathcal{L}}{\partial x^\mu}-\frac{d}{d\lambda}\frac{\partial\mathcal{L}}{\partial \dot{x}^\mu}=0,
\end{equation}
or equivalently \cite{Younsi2016a}
\begin{equation}
\ddot{x}^\mu=g^{\mu\nu}\left(\partial_\nu\mathcal{L}-\partial_\alpha g_{\nu\beta}\dot{x}^\alpha\dot{x}^\beta\right),
\end{equation}
where $\dot{x}=dx/d\lambda$ and
\begin{equation}
\mathcal{L}\equiv p_\mu \dot{x}^\mu-\mathcal{H}=\frac{1}{2}g_{\mu\nu}\dot{x}^\mu\dot{x}^\nu-\frac{1}{2}\omega_p^2.
\end{equation}
There are three integral constants of motion. The stationary and axisymmetric nature of the spacetime implies the conservation of energy $E$ and angular momentum $\Phi$,
\begin{equation}\label{Eq: E and Phi}
E\equiv-p_t,\;\;\;\Phi\equiv p_\phi.
\end{equation}
The last one arises from the Hamiltonian condition $\mathcal{H}=0$.
Carter constant does not exist generally in the presence of a plasma. 
In order to calculate the shadow of Kerr black hole in a plasma, one has to 
solve the equations of motion numerically.

Before carrying out the numerical computation, we now analyze the motion of light rays, which not only serves as a guide for the implementation of the numerical computation, but also  helps in understanding the result of the computation.

Following Ref.[\refcite{Cunha:2016bjh}], we use the effective potential to find the possible region for light propagation. The Hamiltonian condition $\mathcal{H}=0$ reads 
\begin{equation}
\omega_p^2+g^{rr}p^2_r+g^{\theta\theta}p^2_\theta+g^{tt}p^2_t+g^{\phi\phi}p^2_\phi+2g^{t\phi}p_tp_\phi=0.
\end{equation}
Then, a positive definite term can be picked out
\begin{equation}\label{Eq: Kinetic energy}
T\equiv g^{rr}p^2_r+g^{\theta\theta}p^2_\theta\geq0,
\end{equation}
and the Hamiltonian condition could also be written in the form
\begin{equation}\label{Eq: 2 H = T + V}
2\mathcal{H}=T+V=0.
\end{equation}
Therefore, one can identify the problem with a mechanical system with kinetic energy $T$, potential energy $V$ and  
vanishing total energy. Due to the positive definite property of $T$, 
\begin{equation}\label{Eq: effective potential}
V=\omega_p^2+g^{tt}p^2_t+g^{\phi\phi}p^2_\phi+2g^{t\phi}p_tp_\phi\leq0.
\end{equation}
This inequality defines the allowed region for light propagation in the $(r,\theta)$-space. 
Substituting Eq.(\ref{Eq: E and Phi}) into (\ref{Eq: effective potential}), we 
obtain
\begin{equation}\label{Eq: potential equation}
g_{tt}\Phi^2+2g_{t\phi}E\Phi+g_{\phi\phi}E^2-\bar{g}\;\omega_p^2\geq0,
\end{equation}
where we have used the following identities for Kerr spacetime, $g^{rr}g_{rr}=g^{\theta\theta}g_{\theta\theta}=1,\;g^{tt}=-g_{\phi\phi}\bar{g}^{-1},
\;g^{t\phi}=g_{t\phi}\bar{g}^{-1}$ and $g^{\phi\phi}=-g_{tt}\bar{g}^{-1}$. 
Note that $\bar{g}\equiv g^2_{t\phi}-g_{tt}g_{\phi\phi}$ is positive definite 
outside the event horizon.

When $g_{tt}\neq0$, this inequality can be written in the form
\begin{equation}\label{Eq: potential condition}
g_{tt}(\Phi-\lambda_+E)(\Phi-\lambda_-E)\geq0,
\end{equation}
where
\begin{equation}\label{Eq: h_pm}
\lambda_{\pm}\equiv-\frac{g_{t\phi}}{g_{tt}}\pm
\frac{1}{g_{tt}}\sqrt{\bar{g}\left(1+\frac{g_{tt}
		\omega^2_p}{E^2}\right)}.
\end{equation}
It was shown in Ref.[\refcite{Cunha:2016bjh}] that functions $\lambda_{\pm}(r,\theta)$ are useful in determining the boundary of the allowed region in the $(r,\theta)$ space, which is beyond the scope of this paper. Here, we shall show that they are also useful in illustrating how the light rays are  refracted by plasma  near a Kerr black hole.

The existence and value of $\lambda_\pm(r,\theta)$ are dependent on  the conserved energy $E$, which implies that the allowed region would vary according to the photon frequency. Obviously, from Eq.(\ref{Eq: h_pm}) in the limit $E\rightarrow\infty$, 
$\lambda_{\pm}(r,\theta)$ tend to be independent of $\omega_p$. Therefore, it is concluded that the influence of plasma on trajectories of photons with high frequency is negligible. 

The necessary conditions for light propagation inside and outside the ergoregion are different. Inside the ergoregion, by definition, $g_{tt}>0$  and functions $\lambda_{\pm}(r,\theta)$ are real for arbitrary values of $E$. Since $\lambda_+>\lambda_-$, inequality (\ref{Eq: potential condition}) reduces to
\begin{equation}
\Phi>\lambda_+E\;\;\mathrm{or}\;\;\Phi<\lambda_-E,
\end{equation}
which is the necessary condition for light propagation inside the ergoregion. 
On the other hand, outside the ergoregion, $g_{tt}<0$ and functions $\lambda_\pm(r,\theta)$ in Eq.(\ref{Eq: h_pm}) can be rewritten as
\begin{equation}
\lambda_{\pm}=-\frac{g_{t\phi}}{g_{tt}}\pm\frac{1}{g_{tt}}\sqrt{\bar{g}\left(1-\frac{\omega^2_p}{\omega^2}\right)},
\end{equation}
where $\omega$ denotes the photon's frequency and is defined as $\omega\equiv E/\sqrt{-g_{tt}}$. Clearly, the photon motion is possible only when $\omega^2>\omega^2_p$, and the allowed region is defined by $\lambda_-E<\Phi<\lambda
_+E$. When $\omega^2<\omega^2_p$, the photon motion is forbidden, because $\lambda_\pm(r,\theta)$ are now complex functions and the effective potential in Eq.(\ref{Eq: effective potential}) is positive definite outside the ergoreigion.

\section{Ray-tracing algorithm}
\label{sec:RTA}
We shall employ the ray-tracing algorithm to compute the shadow image of a Kerr black hole surrounded by a plasma. To this end,  we assume photons are emitted from the observer's image plane and scattered by the black hole, then according to the principle of light reversibility, the relevant geodesics of light are obtained.

The initial position of each photon is specified by the observer's location. Suppose the observer is far away from the black hole, where the spacetime is well described by Minkowski metric. The location of the observer is denoted as $(r_{o},\theta_o,\phi_o)$ in Boyer-Lindquist coordinates. It is useful to define two sets of Cartesian coordinates $(x,y,z)$ and $(x',y',z')$ centered at the locations of observer and black hole, respectively \cite{Johannsen:2010ru}. 
The $z$-axis is oriented along the radial direction  from the black hole center to the observer.
As such, the image plane $(x,y)$ is perpendicular to $z$-axis, with $x$-axis parallel to $y'$-axis and $y$-axis on the $(z,z')$ plane. 
Therefore, we can transform the coordinates $(x,y,z)$ into $(x',y',z')$ according to the following relationship
\begin{equation}\label{Eq: initial condition posi}
\begin{aligned}
&x'=r_o\sin\theta_o+z\sin\theta_o-y\cos\theta_o,\\&y'=x,\\&z'=r_o\cos\theta_o+z\cos\theta_o+y\sin\theta_o.
\end{aligned}
\end{equation}
Here, without loss of generality, we have fixed $\phi_o=0$.
The transformation between the black hole's Cartesian and  spherical coordinates is given as usual
\begin{equation}\label{Eq: Cord TM}
\begin{aligned}
&r=\sqrt{x'^2+y'^2+z'^2},\\&\theta=\arccos\frac{z'}{r},\\&\phi=\arctan\frac{y'}{x'}.
\end{aligned}
\end{equation}
Since photons are emitted from the observer's image plane, their initial position is $(x,y,0)$ in the observer's coordinates, and the 3-momentum gives $\bm{p}=p_0\bm{e}_z$, where $p_0$ is the photon frequency measured by the observer. Then, by combining Eqs.(\ref{Eq: initial condition posi}) and (\ref{Eq: Cord TM}),  the corresponding initial values of $(r,\theta,\phi)$ for a given photon are obtained. Note that the initial value of $t$ does not affect the light rays, so it is safe to set $t=0$ initially.

On the other hand, the initial momenta $(\dot{r},\dot{\theta},\dot{\phi})$ for the given photon are determined by
\begin{equation}
\dot{r}=-p_0\cos\theta\cos\theta_o-p_0\sin\theta\sin\theta_o\cos\phi,
\end{equation}
\begin{equation}
\dot{\theta}=-\frac{p_0}{r}\left[\sin\theta_o\cos\theta\cos\phi-\sin\theta\cos\theta_o\right],
\end{equation}
\begin{equation}
\dot{\phi}=\frac{p_0}{r\sin\theta}\sin\theta_o\sin\phi.
\end{equation}
And the value of $\dot{t}$ is computed from the Hamiltonian condition, $\mathcal{H}=0$, i.e.,
\begin{equation}
\dot{t}=\beta+\sqrt{\beta^2+\gamma},
\end{equation}
where the quantities $\beta$ and $\gamma$ are defined as
\begin{equation}
\beta\equiv-\frac{g_{ti}\dot{x}^i}{g_{tt}}
\end{equation}
and
\begin{equation}
\gamma\equiv-\frac{g_{ij}\dot{x}^i\dot{x}^j+\omega_p^2}{g_{tt}},
\end{equation}
respectively.

With the initial conditions given above, we integrate the second-order differential  equations (\ref{Eq: EL eqs}) backwards using a fourth-order Runge-Kutta integrator.  The integration goes on until photons fall into the black hole or escape to infinity. It should be noted that $|\dot{t}|$ tends to diverge near the event horizon and one needs a small step size for the affine parameter $\lambda$ to maintain numerical precision. In order to improve the efficiency, a variable step strategy is applied. Starting from an initial value of $\delta\lambda$, at each integrating step,  the step size is renewed as \cite{Psaltis:2010ww}
\begin{equation}\label{Eq: variable stepsize}
\delta\lambda=f\min\left\lbrace \biggr|\frac{t}{\dot{t}}\biggr|,\biggr|\frac{r}{\dot{r}}\biggr|,\biggr|\frac{\theta}{\dot{\theta}}\biggr|,\biggr|\frac{\phi}{\dot{\phi}}\biggr| \right\rbrace ,
\end{equation}
where $f$ is a fixed coefficient.  In the running of our codes, we find that the strategy works well and the step size in Eq.(\ref{Eq: variable stepsize})  takes a small value automatically near the event horizon.

\section{Results}
\label{sec:results}
\subsection{The Kerr shadow in a plasma}
In the above ray-tracing algorithm, there are only two possible final states of photons emitted from the observer's screen: one possibility is that the photon is absorbed by the black hole and the other is that the photon is scattered to infinity. Then, the shadow of a black hole is a region in observer's image plane from which the emitted photons will finally fall into the black hole. 

In vacuum, the region that allows the photon to propagate does not depend on the photon's frequency \cite{Cunha:2016bjh}. It follows that different values of $p_0$ will generate the same shadow images, and without loss of generality one may choose $p_0=1$ in the numerical integration of null geodesic equations. 
However, when a plasma is taken into  consideration, the region that allows photons to propagate depends on the photon's frequency. Therefore, it is expected that the black hole shadow will take on a  different look from the vacuum case in the way that depends on the photon frequency, and at the high-frequency limit, the influence of plasma on the motion of photon is negligible, then,  the black hole shadow should be well approximated by those in vacuum. These are confirmed by our numerical results.
 
\begin{figure*}
	\centering
	\includegraphics[width=0.32\textwidth,angle=0]{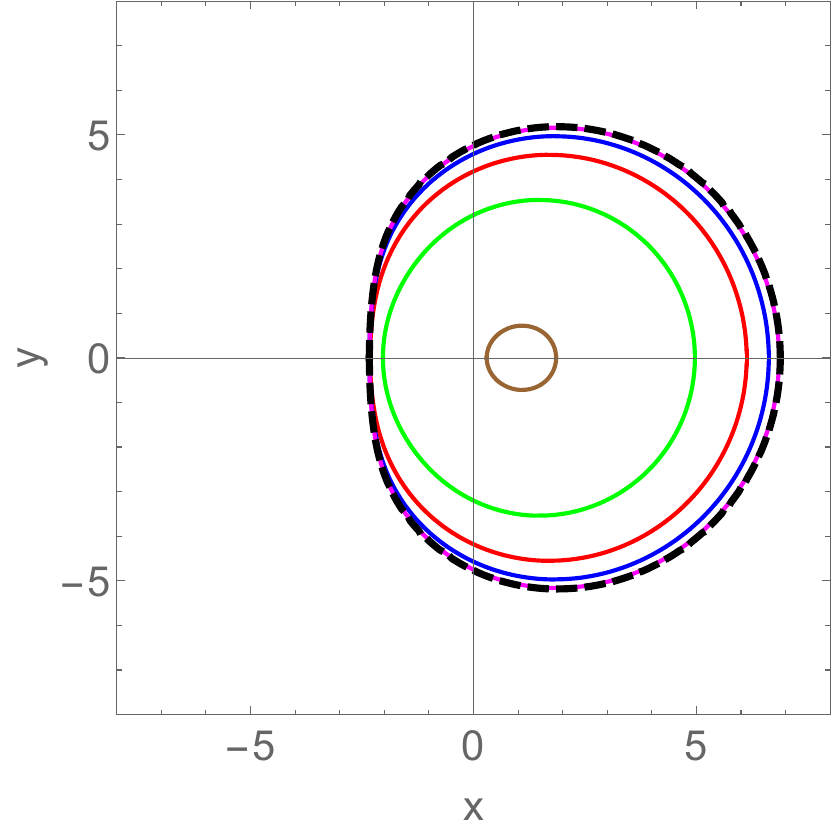}
	\includegraphics[width=0.32\textwidth,angle=0]{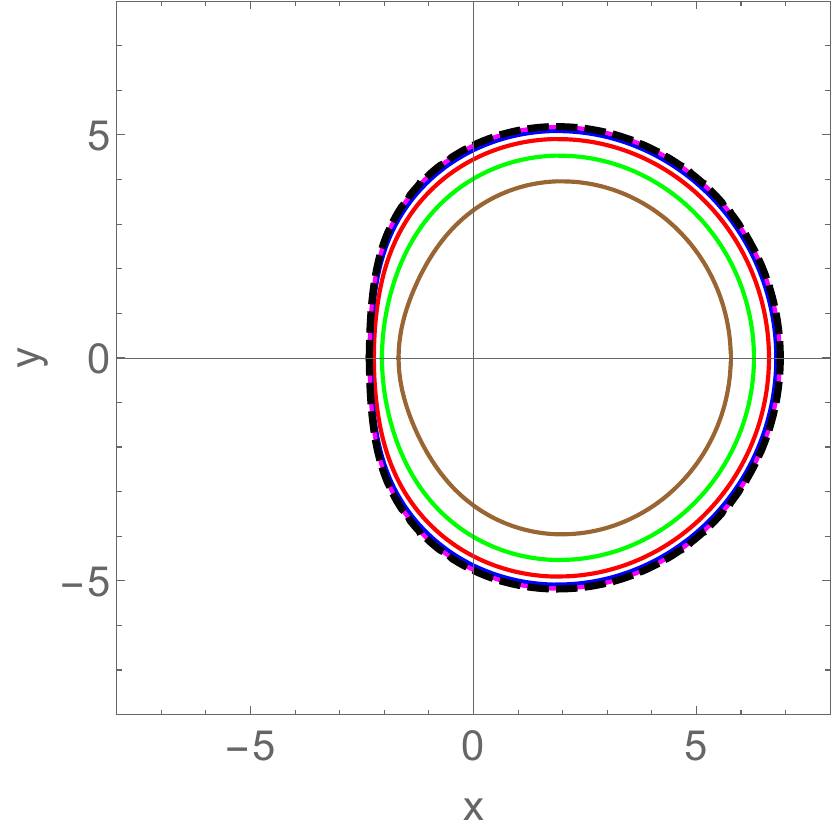}
	\includegraphics[width=0.32\textwidth,angle=0]{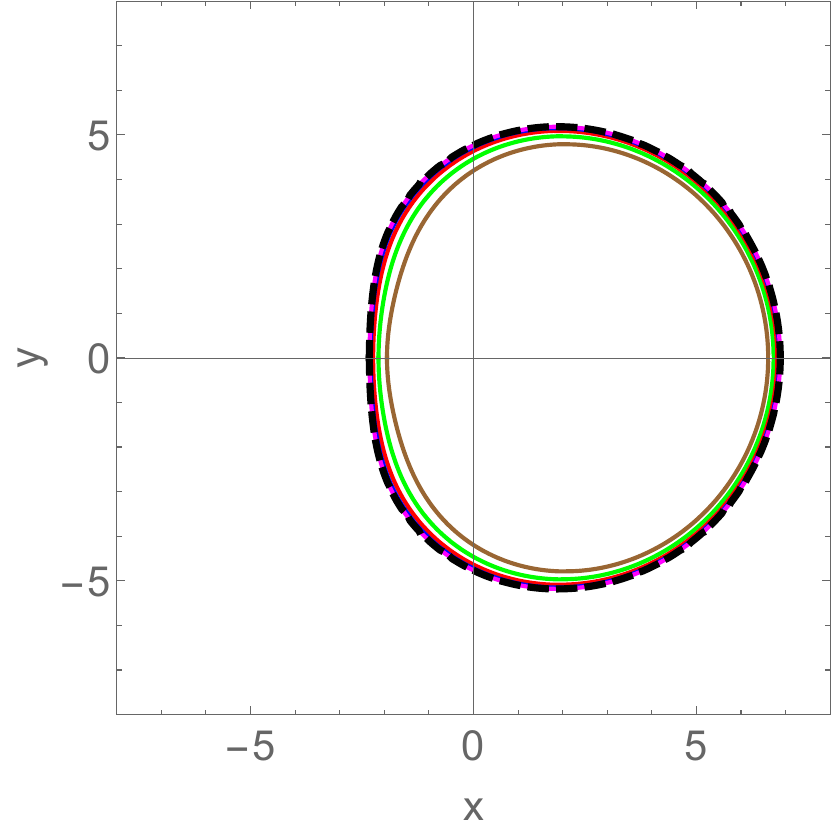}
	\caption{Shadows of Kerr black hole ($a=0.99$) in the presence of a plasma in the power-law model for different values of $p_0$: $p_0=3.0$ (magenta), $1.0$ (blue), $0.6$ (red), $0.4$ (green), $0.3$ (brown), from outermost to innermost. Three different values of parameter $h$ in the model are considered, i.e., $h=1$ (left), $2$ (middle), and $3$ (right). The density parameter of the plasma is $k_0=1$. For the sake of comparison, we also plot the Kerr black hole shadow (the black dashed curve in the pictures) without plasma.}
	\label{Fig: p0_shadow}
\end{figure*}

\begin{figure*}
	\centering
	\includegraphics[width=0.8\textwidth,angle=0]{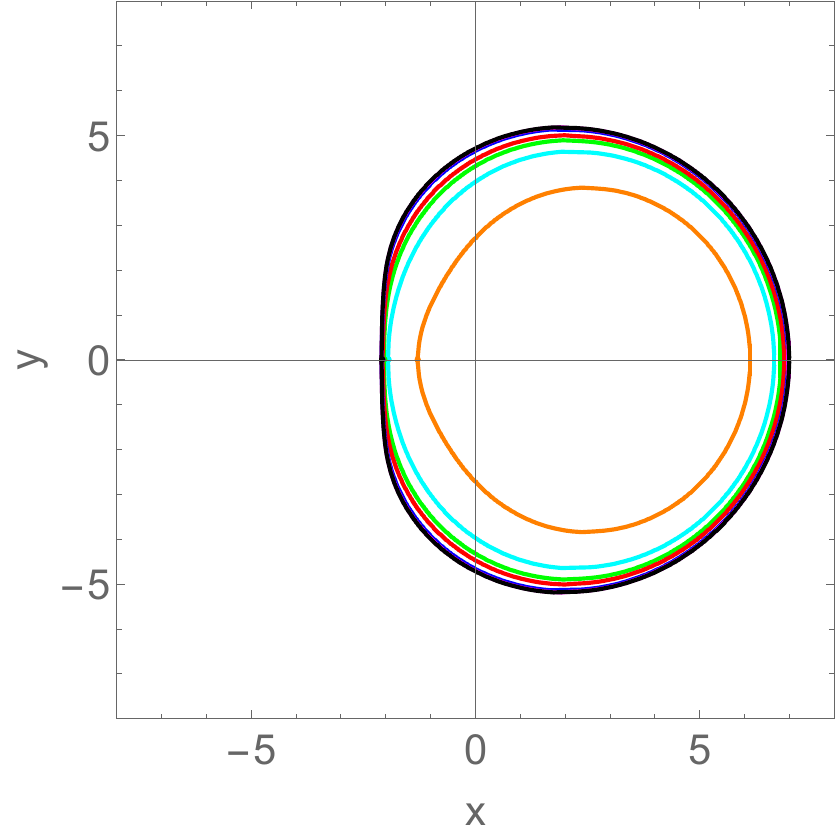}
	\caption{Shadows of Kerr black hole ($a=0.998$) in a plasma of exponential model for different values of $p_0$: $p0=3.0,2.0,1.0,0.5,0.4,0.3,0.2$, from the outermost to the innermost. The two parameters of plasma are chosen as $k_0=1$ and $r_0=20M$, respectively.}
	\label{Fig: p0_shadow_exponential}
\end{figure*}

Figure \ref{Fig: p0_shadow} shows the dependence of Kerr black hole shadows on $p_0$ in the power-law model.   It is observed that when $p_0\gtrsim3$ the shadow of black hole in plasma is indistinguishable from that in vacuum. As $p_0$ goes to small, the influence of plasma appears: the shadow shrinks as $p_0$ decreases. When $p_0$ is decreased close to some critical values, the Kerr shadow becomes a point and will disappear in the observer's screen. Similar results can be also obtained for the exponential model, see figure \ref{Fig: p0_shadow_exponential}.

\begin{figure}
	\centering
	\includegraphics[width=0.8\textwidth,angle=0]{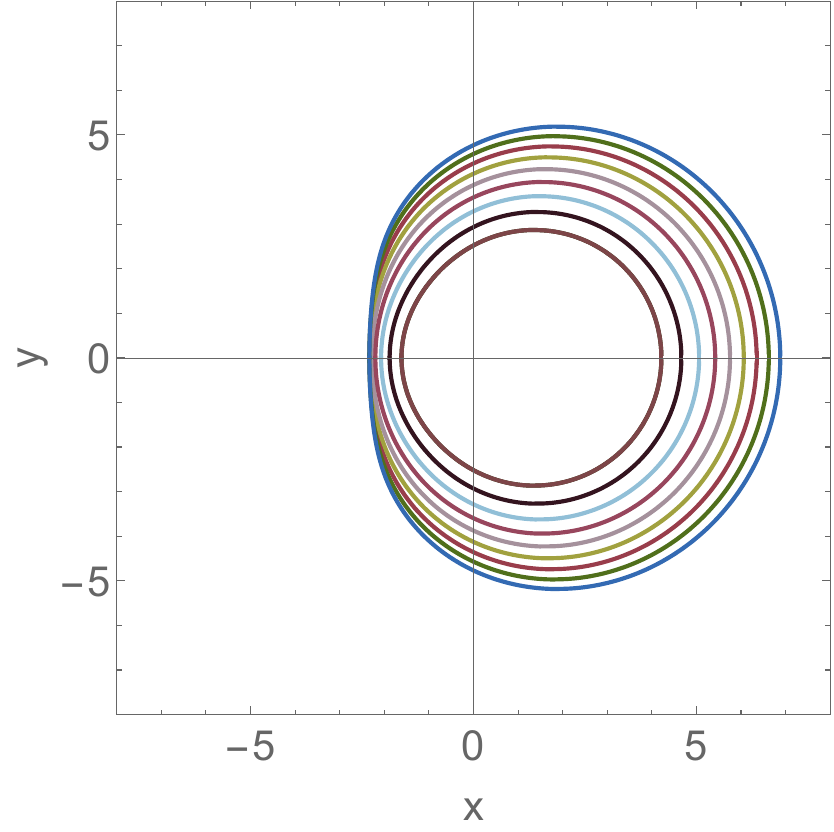}
	\caption{Influence of the density parameter of plasma $k_0$ on the Kerr black hole shadow in the power-law model. $k_0=0,1,\cdots,9$ from outermost to innermost. Here, we choose $p_0=1$, $a=0.99$, and $h=1$.}
	\label{Fig: k0_shadow}
\end{figure}

We note that, in the power-law model, the smaller value of $h$ corresponds to the higher density of the plasma at a given position. Consequently, as shown in figure \ref{Fig: p0_shadow}, for the same value of $p_0$, the Kerr shadow with $h=1$ is smaller than those with $h=2$ and $3$. Another way to change the density of the plasma is to choose different values of $k_0$. Figure \ref{Fig: k0_shadow} shows the dependence of Kerr shadow on $k_0$. Clearly, $k_0=0$ means that no plasma stays outside the black hole and we obtain the Kerr shadow in the vacuum. From figure \ref{Fig: k0_shadow}, when $k_0$ increases, the shadow shrinks again as expected.

\subsection{The shape parameters}
To quantify the effects of plasma on the shape of Kerr black hole shadow, it is useful to introduce several shape parameters, for example, $\left\lbrace x_{\mathrm{max}},\;x_{\mathrm{min}},\;D,\;D_x,\;\sigma\right\rbrace$ \cite{PhysRevLett.115.211102,Johannsen:2010ru}. Here $x_{\mathrm{max}}$ and $x_{\mathrm{min}}$ are the maximum and minimum abscissae of the shadow, respectively; the horizontal displacement can be represented by $D={|x_{\mathrm{max}}+x_{\mathrm{min}}|}/{2}$, the distance of a point on the shadow's edge from the center is $r=\sqrt{(x-D)^2+y^2}$; $\sigma$ gives the deviation from roundness, i.e.,
\begin{equation}\label{Eq: sigma}
\sigma=\sqrt{\frac{1}{2\pi}\int_{0}^{2\pi}\left[r(\alpha)-\bar{r}\right]^2d\alpha},
\end{equation}
where $\bar{r}$ is the average distance, which is give by $\bar{r}=\frac{1}{2\pi}\int_{0}^{2\pi}r(\alpha)d\alpha$. Finally,
\begin{equation}
D_x\equiv|x_{\text{max}}-x_{\text{min}}|
\end{equation}
describes the horizontal size of the black hole shadow. 
Here, we are mainly focussed on the influence of $k_0$ and 
$p_0$ on the shape parameters  $\sigma$ and $D_x$. 

\begin{figure*}
	\centering
	\includegraphics[height=0.25\textheight,width=0.45\textwidth]{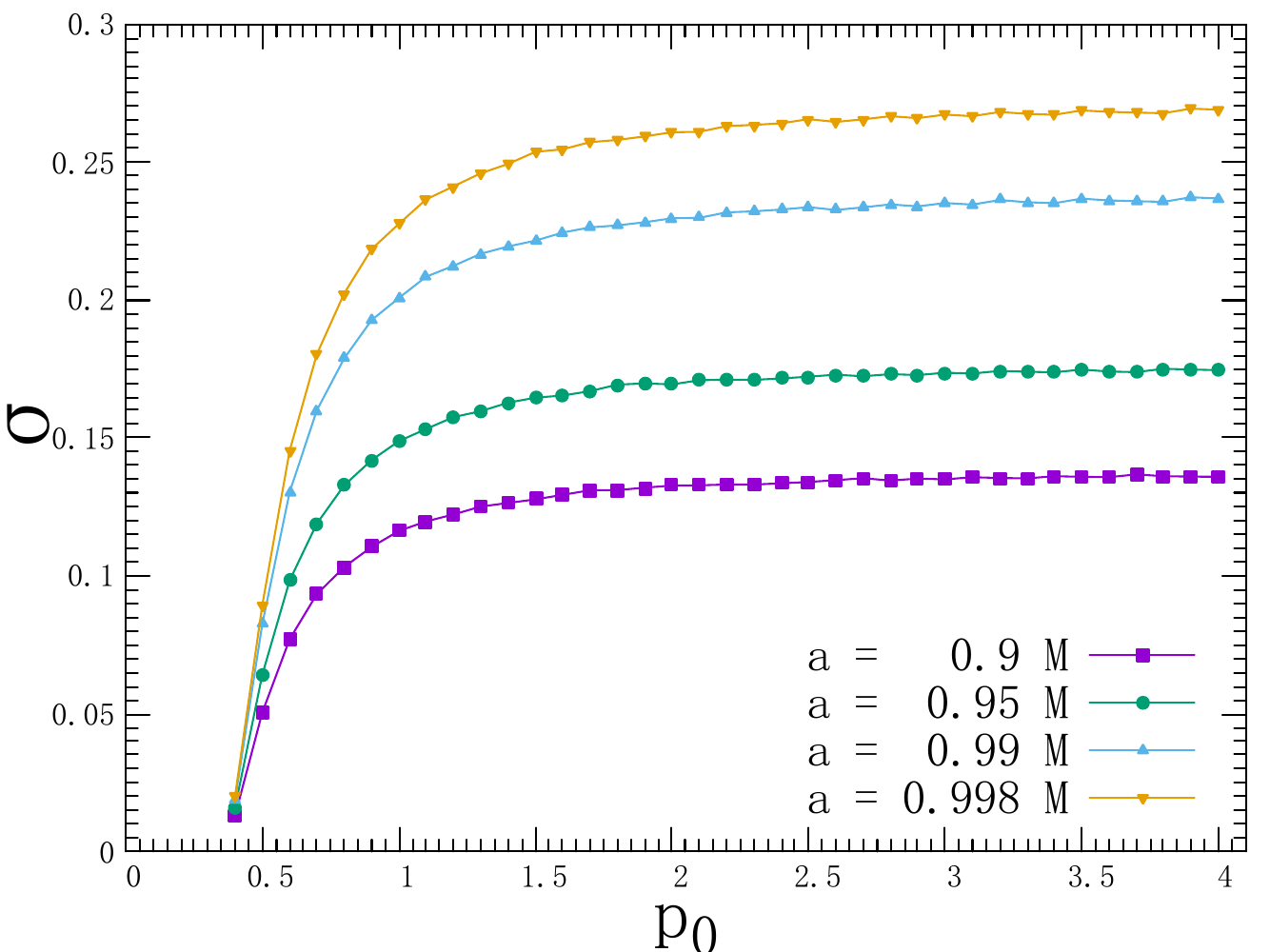}
	\includegraphics[height=0.25\textheight,width=0.45\textwidth]{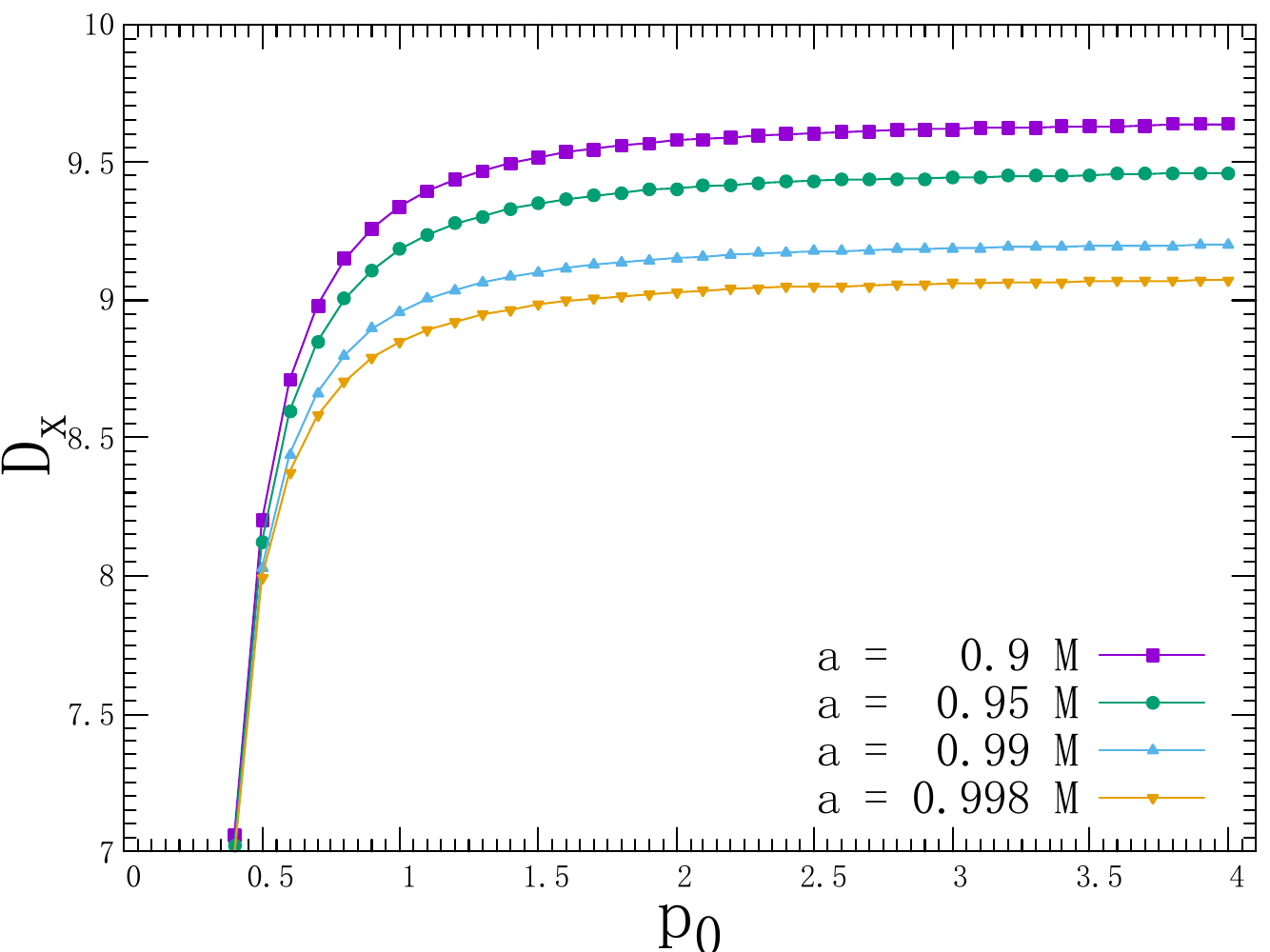}

	\includegraphics[height=0.25\textheight,width=0.45\textwidth]{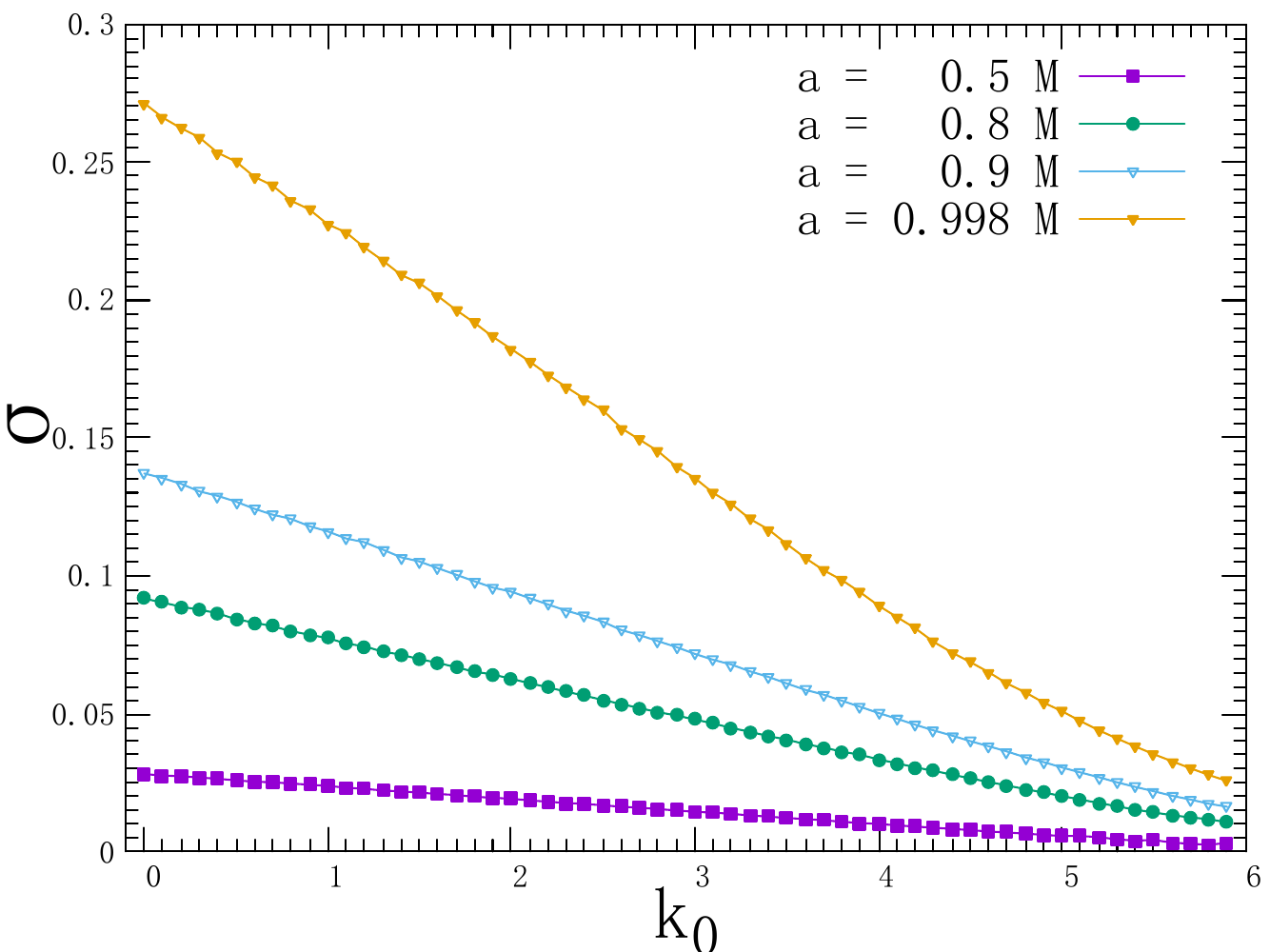}
	\includegraphics[height=0.25\textheight,width=0.45\textwidth]{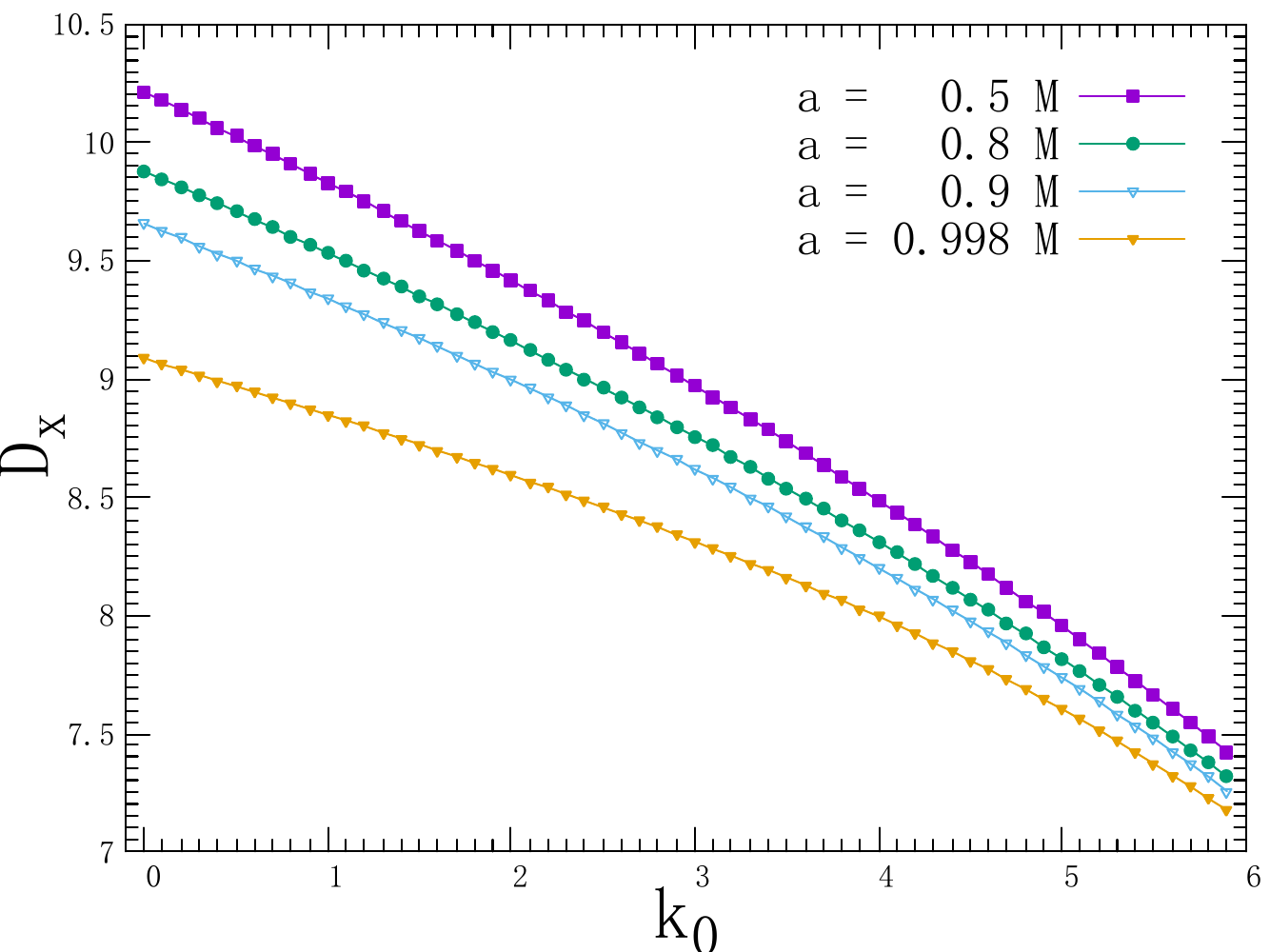}
	\caption{The shape parameters $\sigma$ (left column) and $D_x$ (right column) 
		of the Kerr shadow for various values of $a$, with $h=1$ for the power law 
		distribution of the plasma in Eq.(\ref{Eq: plasma md}). Top panel: The 
		shape parameters as functions of $p_0$, with $k_0=1$. Bottom panel: 
		The shape parameters as functions of $k_0$, with $p_0=1$.}
	\label{Fig: sigma and Dx vs p0}
\end{figure*}

\begin{figure*}
	\centering
	\includegraphics[height=0.25\textheight, width=0.45\textwidth]{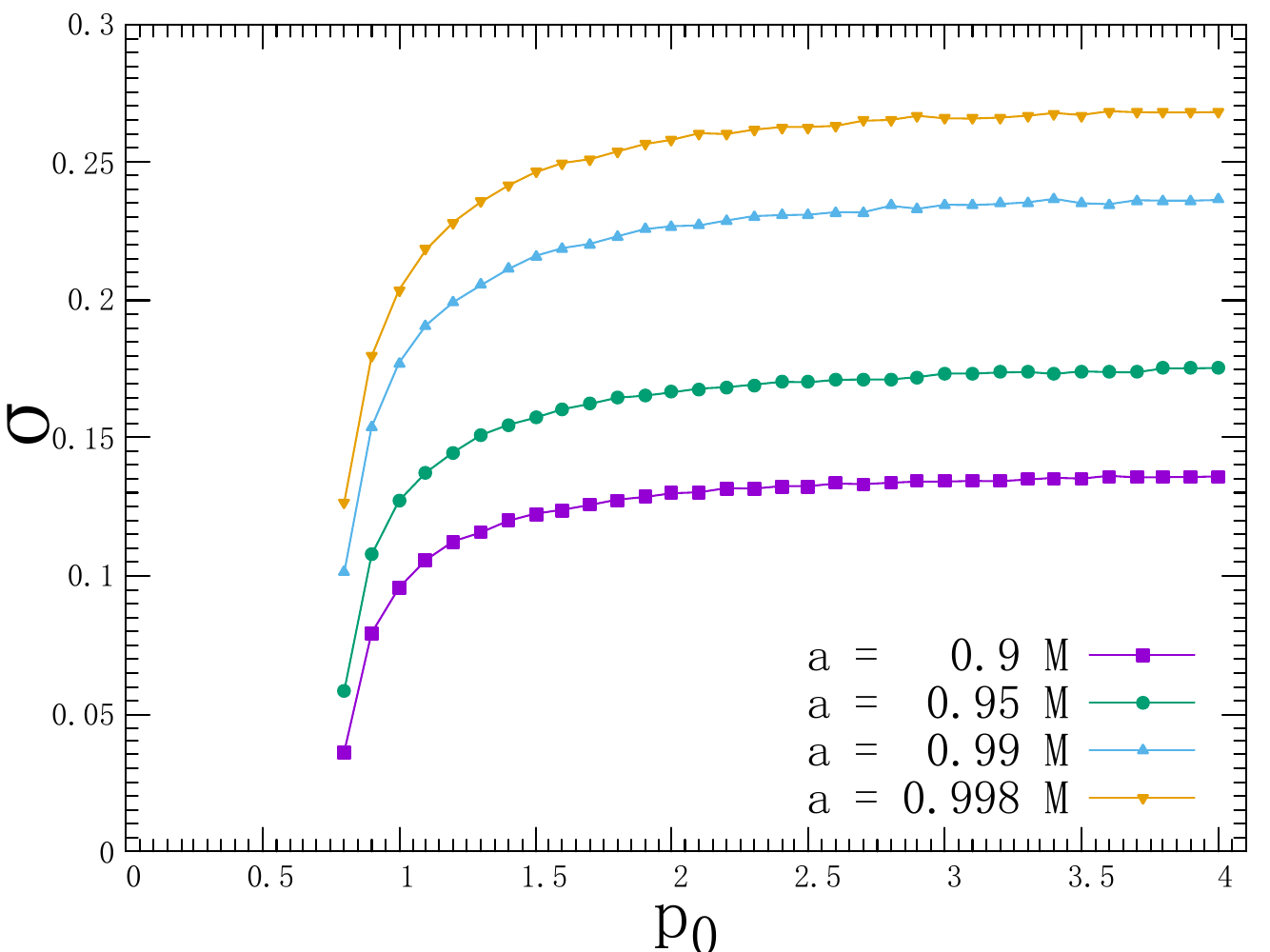}
	\includegraphics[height=0.25\textheight,width=0.45\textwidth]{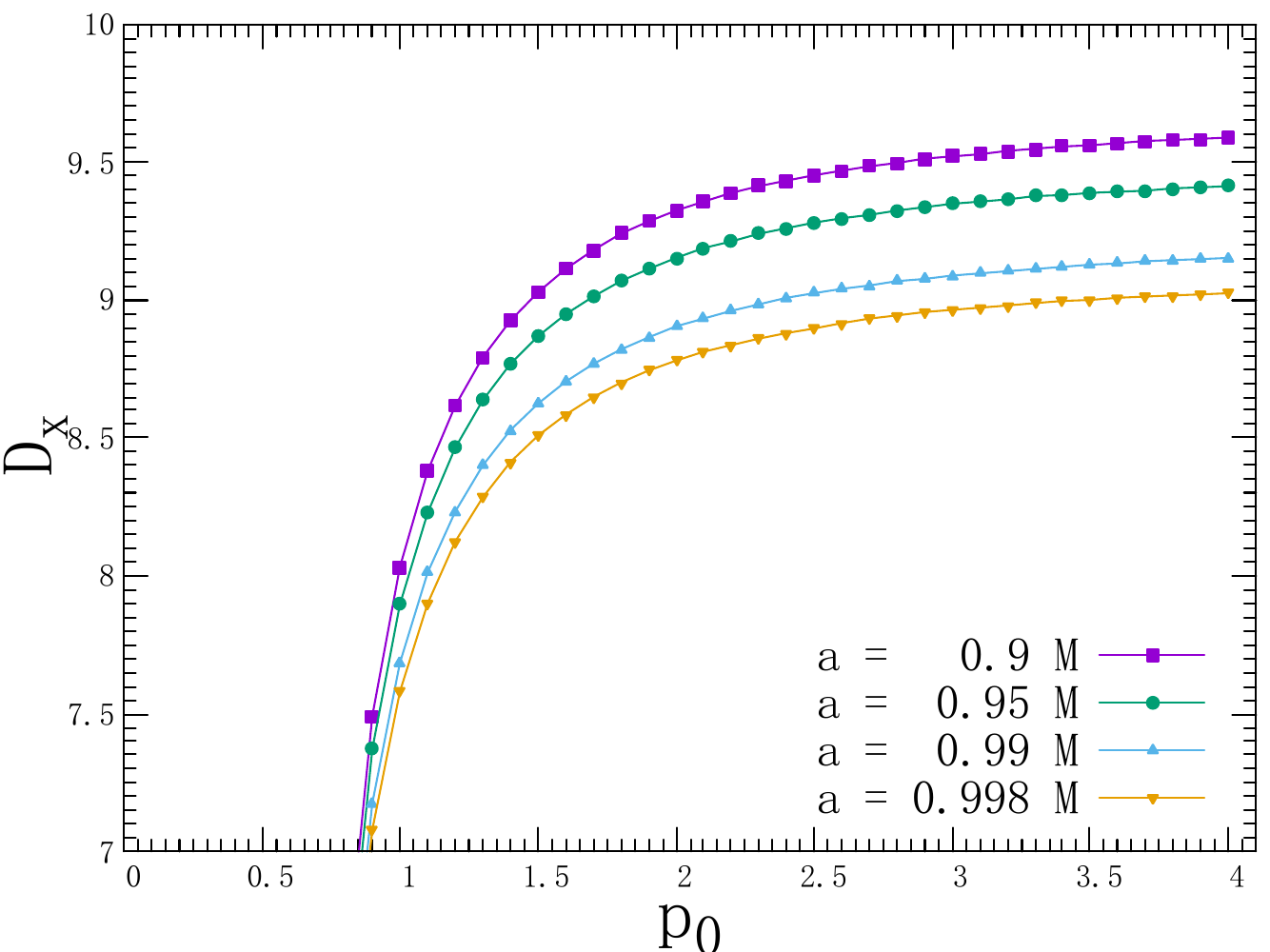}
	\caption{The shape parameters $\sigma$ (left panel) and $D_x$ (right panel) of the Kerr shadow in a plasma with a radial exponential distribution for different values of $a$. The two parameters of plasma are chosen as $k_0=1$ and $r_0=20M$, respectively.}
	\label{fig_sigma_Dx_vs_p0_Exponential}
\end{figure*}

Figure \ref{Fig: sigma and Dx vs p0} shows the dependence of shape 
parameters $\sigma$ and $D_x$ of shadows in power-law model on $p_0$ and model parameter $k_0$, in the top and bottom panels respectively. 
From the top panel, there is a sharp decrease in the shape parameters 
when $p_0$ takes small values, typically for $p_0\leq0.8$. However, 
when $p_0$ is increased to large values, the shape parameter $\sigma$, as well as $D_x$,  approaches slowly to  a constant value which correspond to the value in the vacuum case.
From figure \ref{fig_sigma_Dx_vs_p0_Exponential}, the similar dependence of shape parameters of shadow in exponential model is also found.   
When we consider the case in which $p_0=\text{const}$, as shown 
in the bottom panel of figure \ref{Fig: sigma and Dx vs p0}, the Kerr shadow 
in the vacuum corresponds to $k_0=0$. When $k_0\neq0$, it is found that both 
$\sigma$ and $D_x$ decrease as $k_0$ increases. The Kerr shadow becomes 
more round and smaller in a denser plasma. This is consistent with the result 
in figure \ref{Fig: k0_shadow}.

\subsection{Lensing by Kerr black hole in a plasma}
\begin{figure}
	\centering
	\includegraphics[height=0.3\textheight]{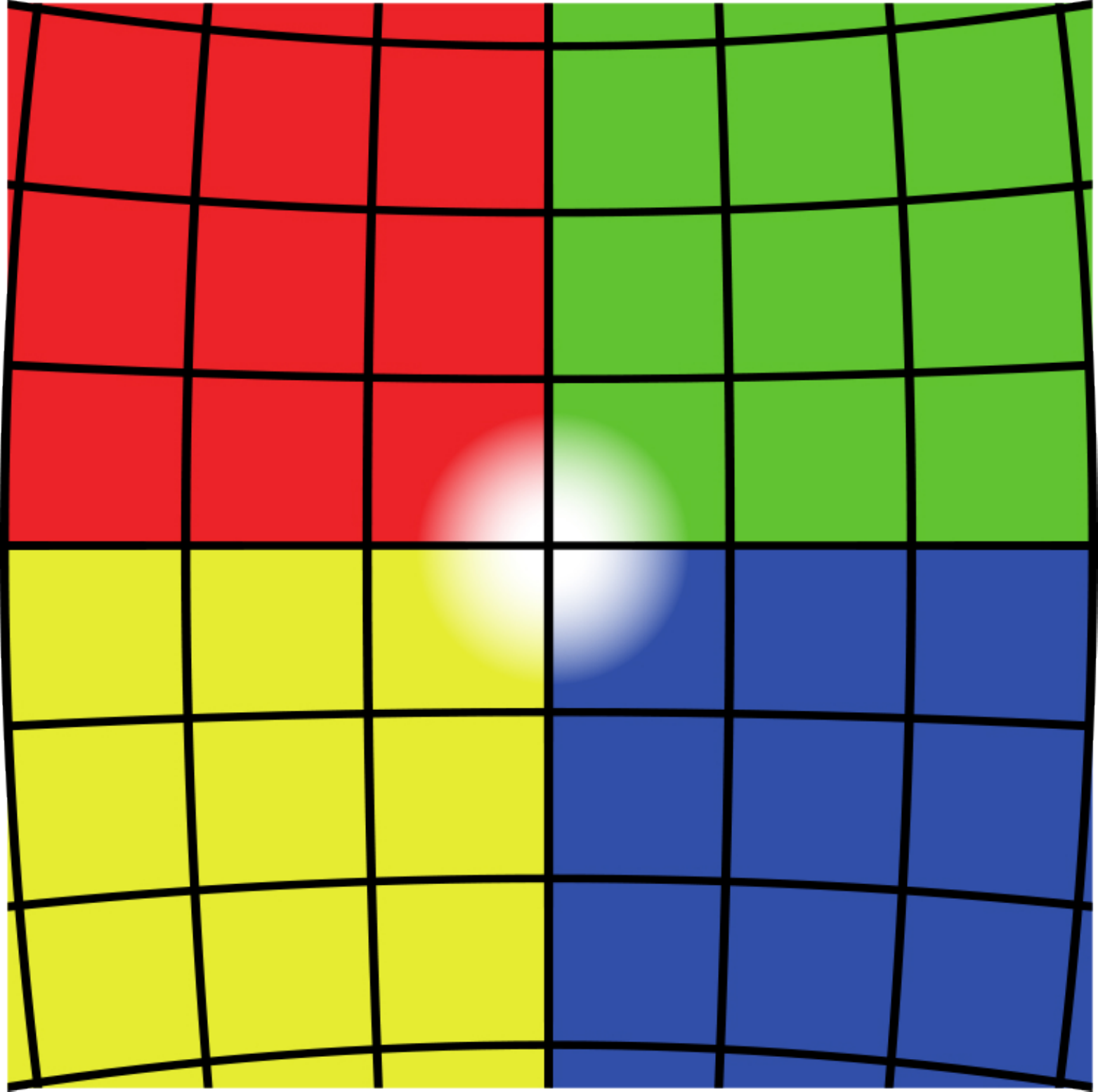}	
	\caption{The observer's field of view when there is no black hole at the center of the celestial sphere.}
	\label{Fig: background}
\end{figure}

In order to demonstrate our numerical results more intuitively, it is convenient to create the  "images" of black holes with different parameters and surrounded by plasma with different density distribution. 
To this end, following Refs. [\refcite{Bohn:2014xxa,PhysRevLett.115.211102,Wang:2017qhh}], we use the celestial sphere as the artificial background. Furthermore, the celestial sphere is divided into four quadrants, and each marked with a different color. Suppose the black hole is located at the center and an observer who looks at the center of the celestial sphere is placed off-center inside the celestial sphere. Then, we assign each point a color in observer's screen according to the final location of the photon that emitted from this point. When the black hole is missing, the observer's field of vision is illustrated in figure \ref{Fig: background}. In this plot, the white reference spot lies at the observation direction, which will form an Einstein ring caused by the gravitational lensing of a black hole. We also include the longitude (vertical) and latitude (horizontal) lines to illustrate the spatial distortion caused by the black hole and plasma. 

\begin{figure}
	\centering
	\includegraphics[height=0.19\textwidth,width=0.19\textwidth,angle=90]{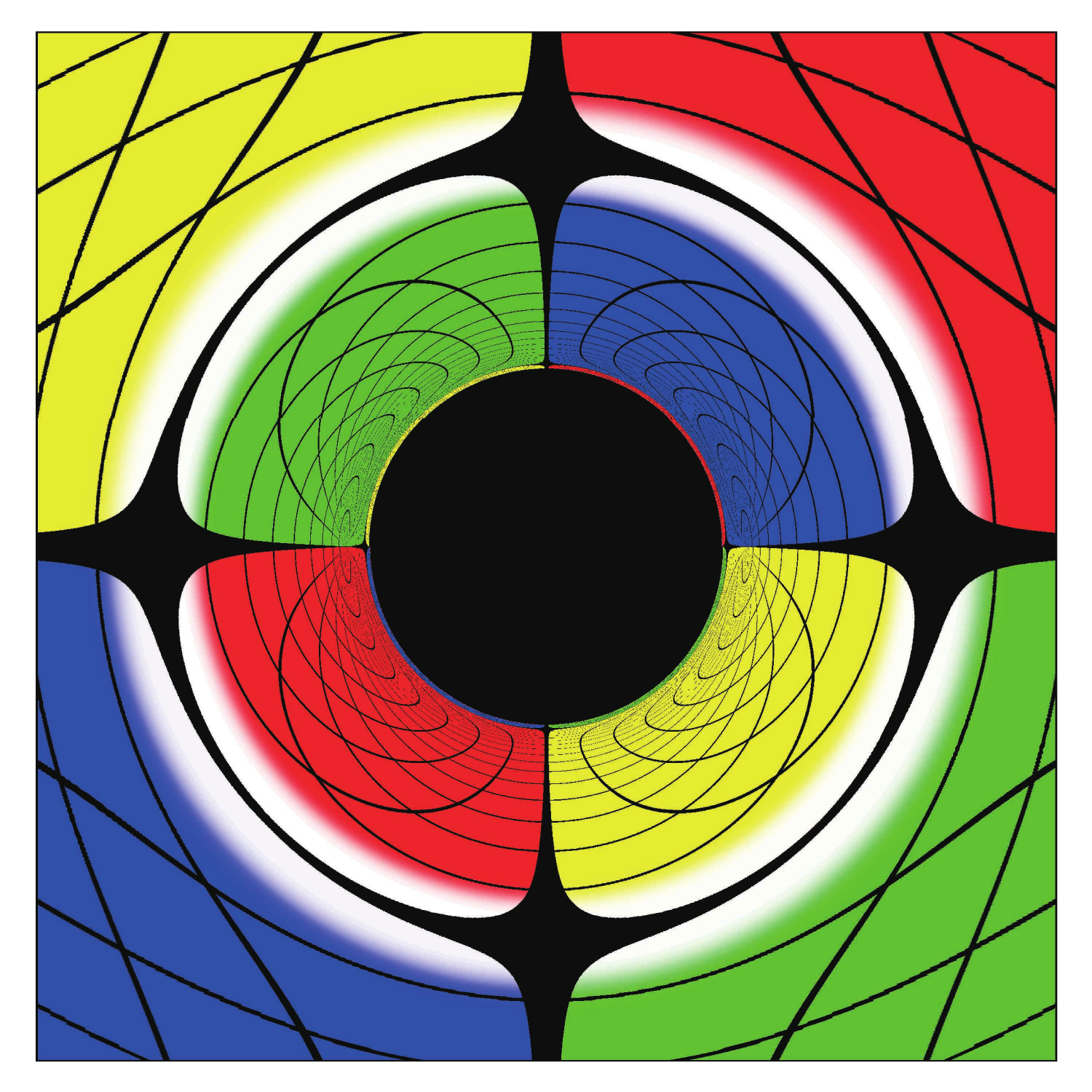}
	\includegraphics[height=0.19\textwidth,width=0.19\textwidth,angle=90]{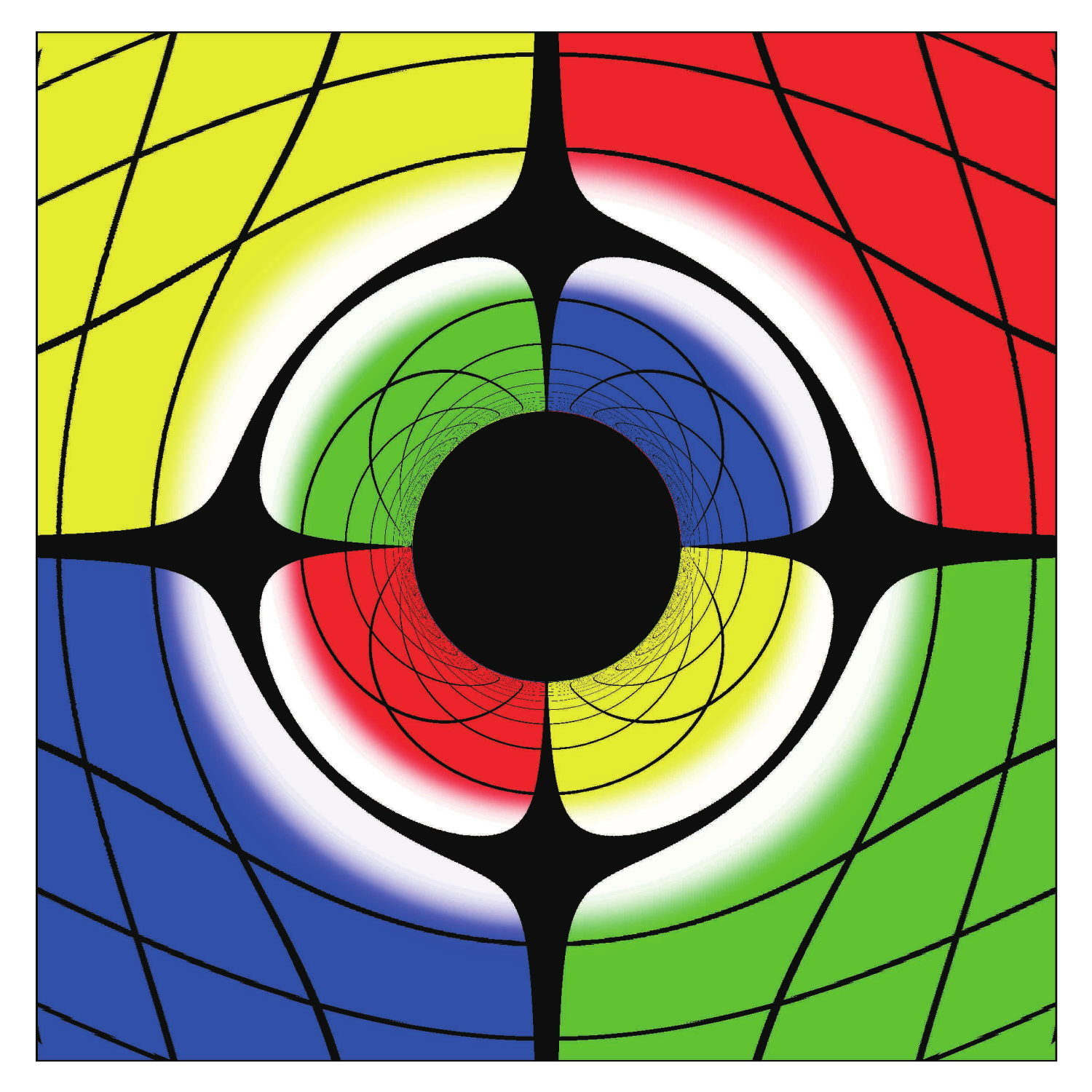}		\includegraphics[height=0.19\textwidth,width=0.19\textwidth,angle=90]{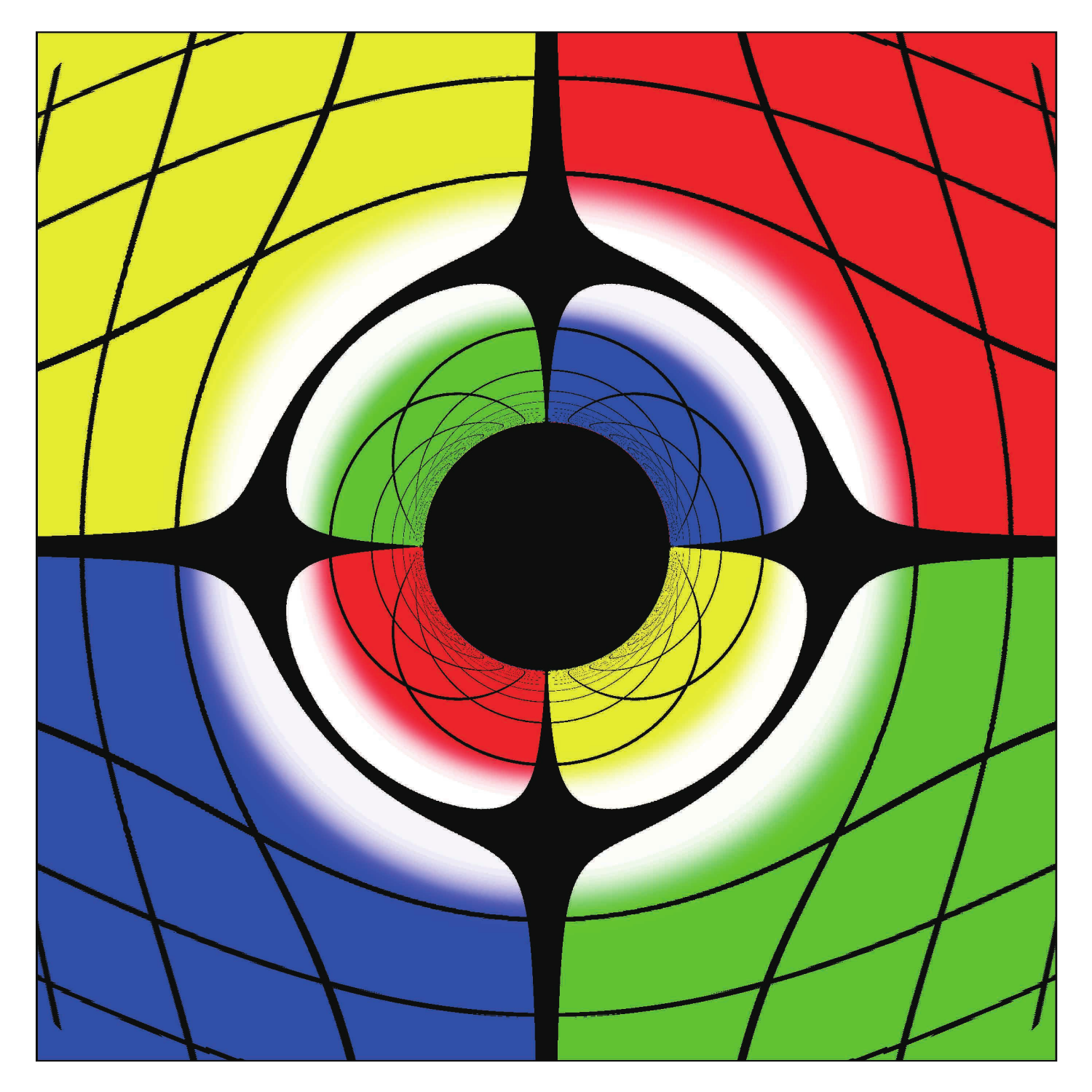}
	\includegraphics[height=0.19\textwidth,width=0.19\textwidth,angle=90]{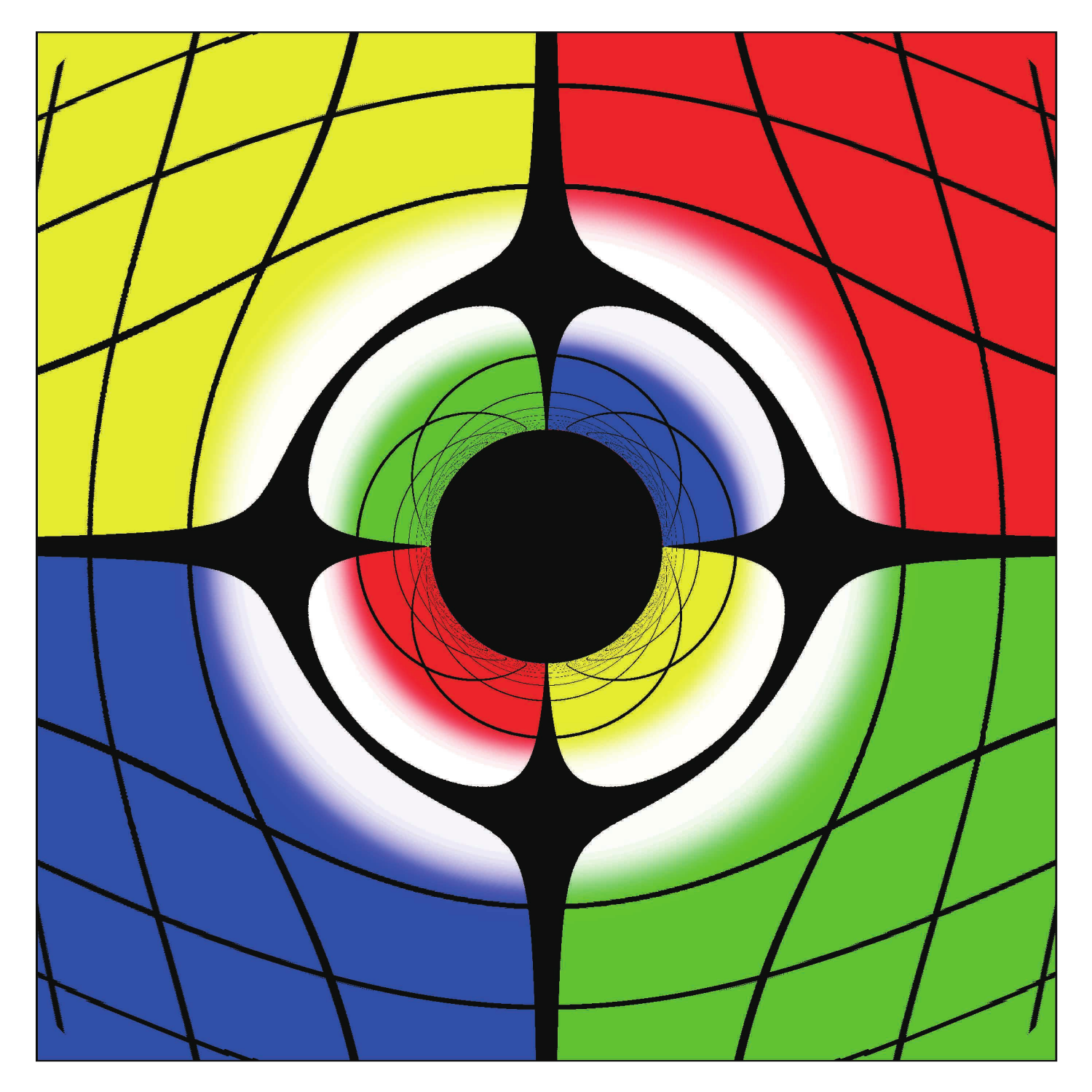}
	\includegraphics[height=0.19\textwidth,width=0.19\textwidth,angle=90]{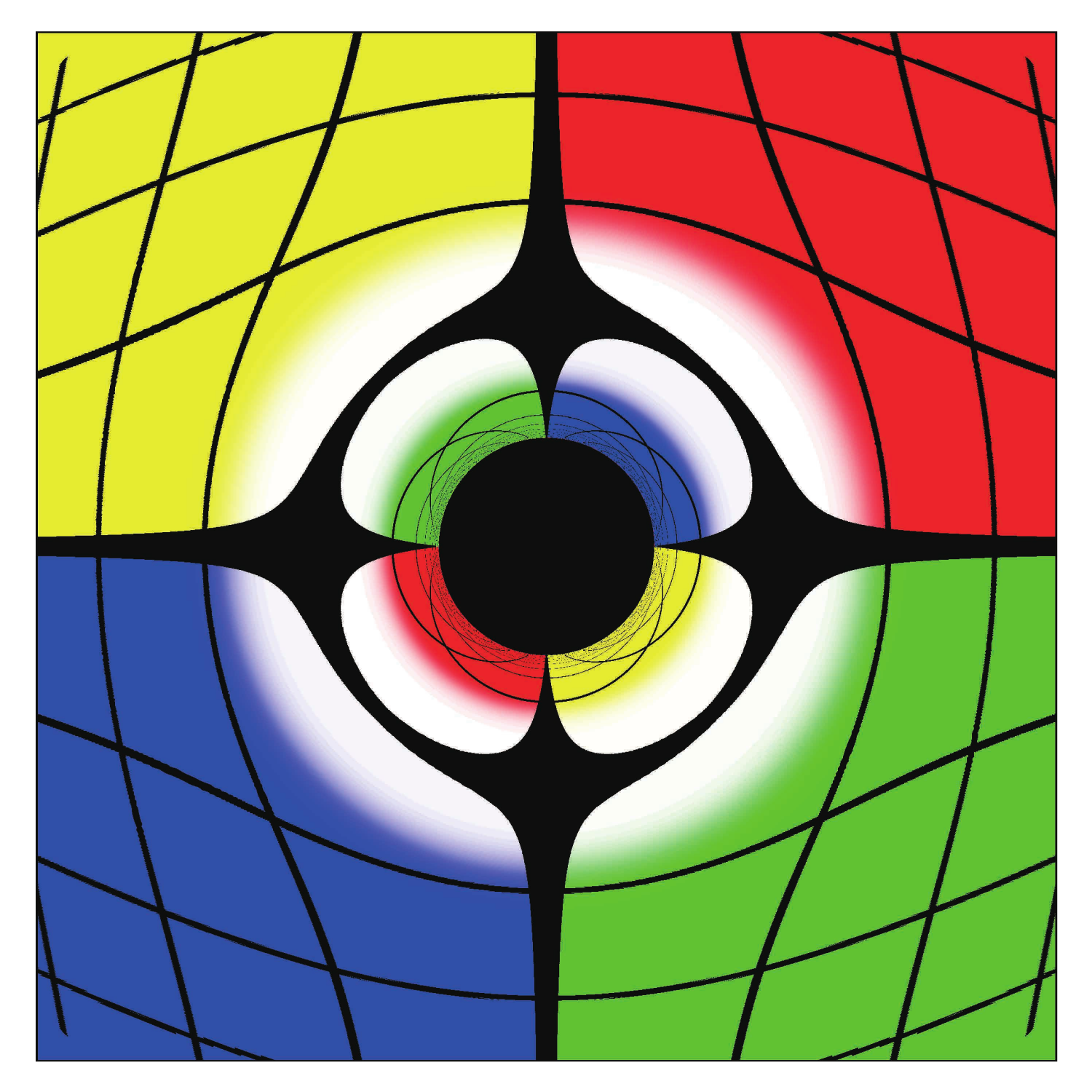}
	
	\includegraphics[height=0.19\textwidth,width=0.19\textwidth,angle=90]{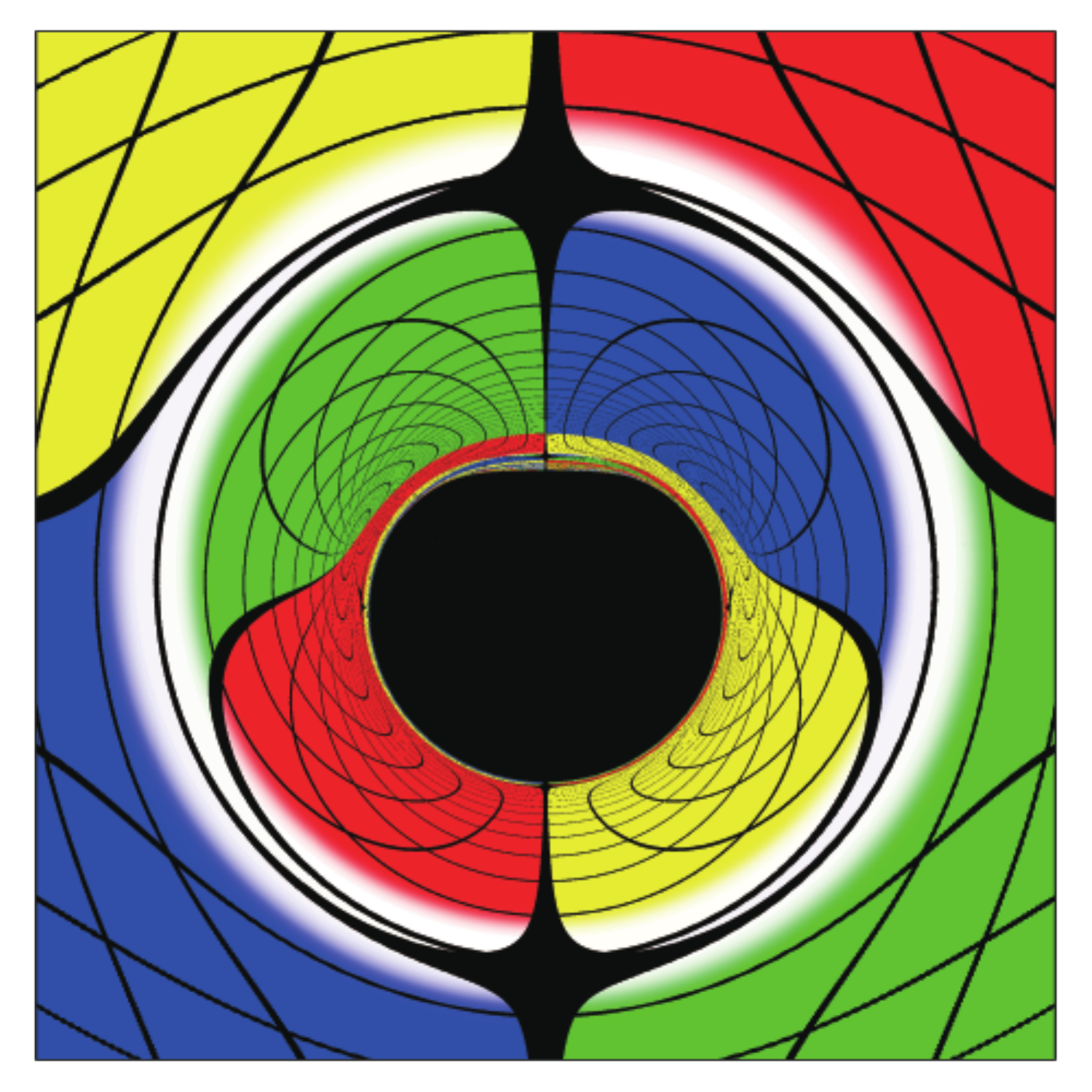}
	\includegraphics[height=0.19\textwidth,width=0.19\textwidth,angle=90]{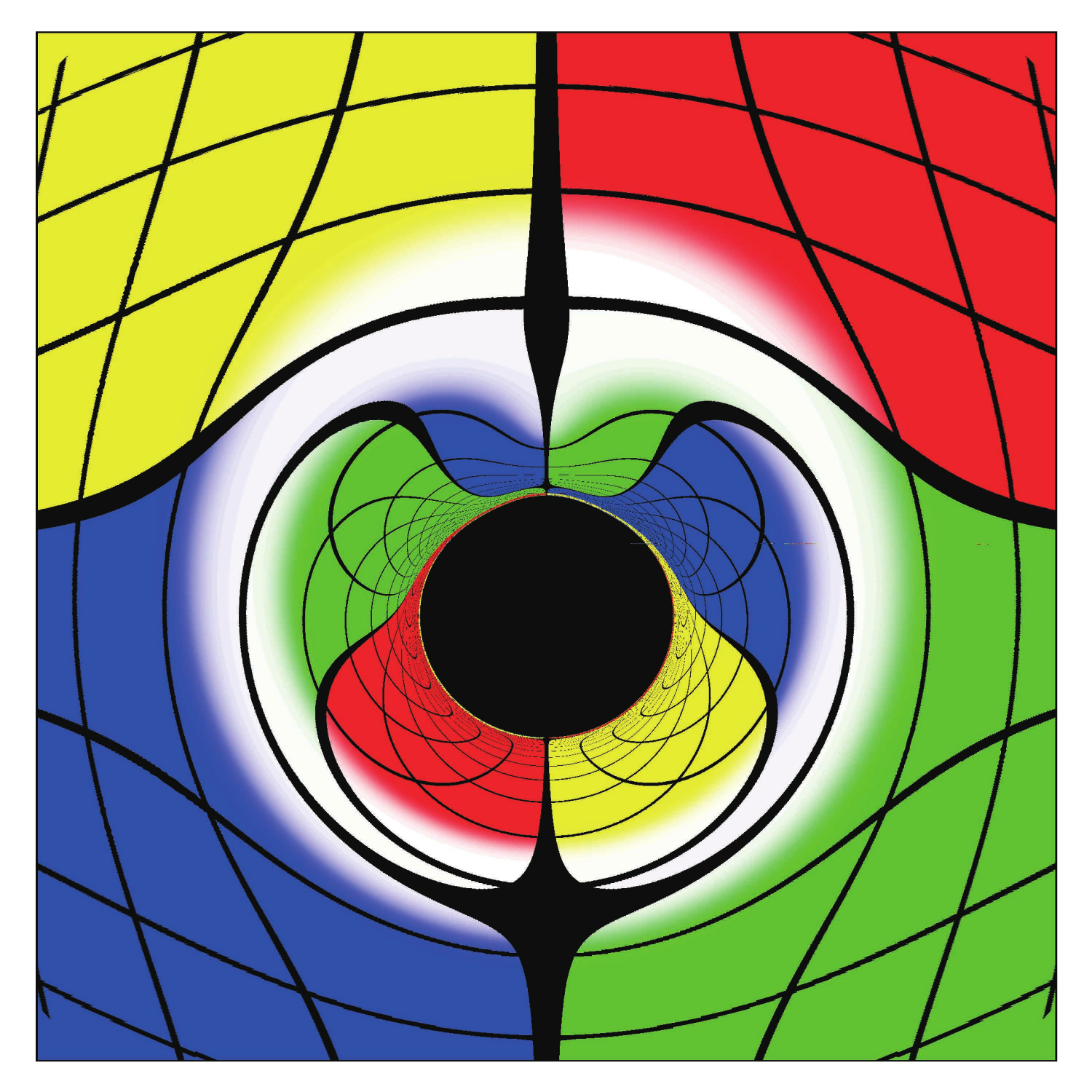}		
	\includegraphics[height=0.19\textwidth,width=0.19\textwidth,angle=90]{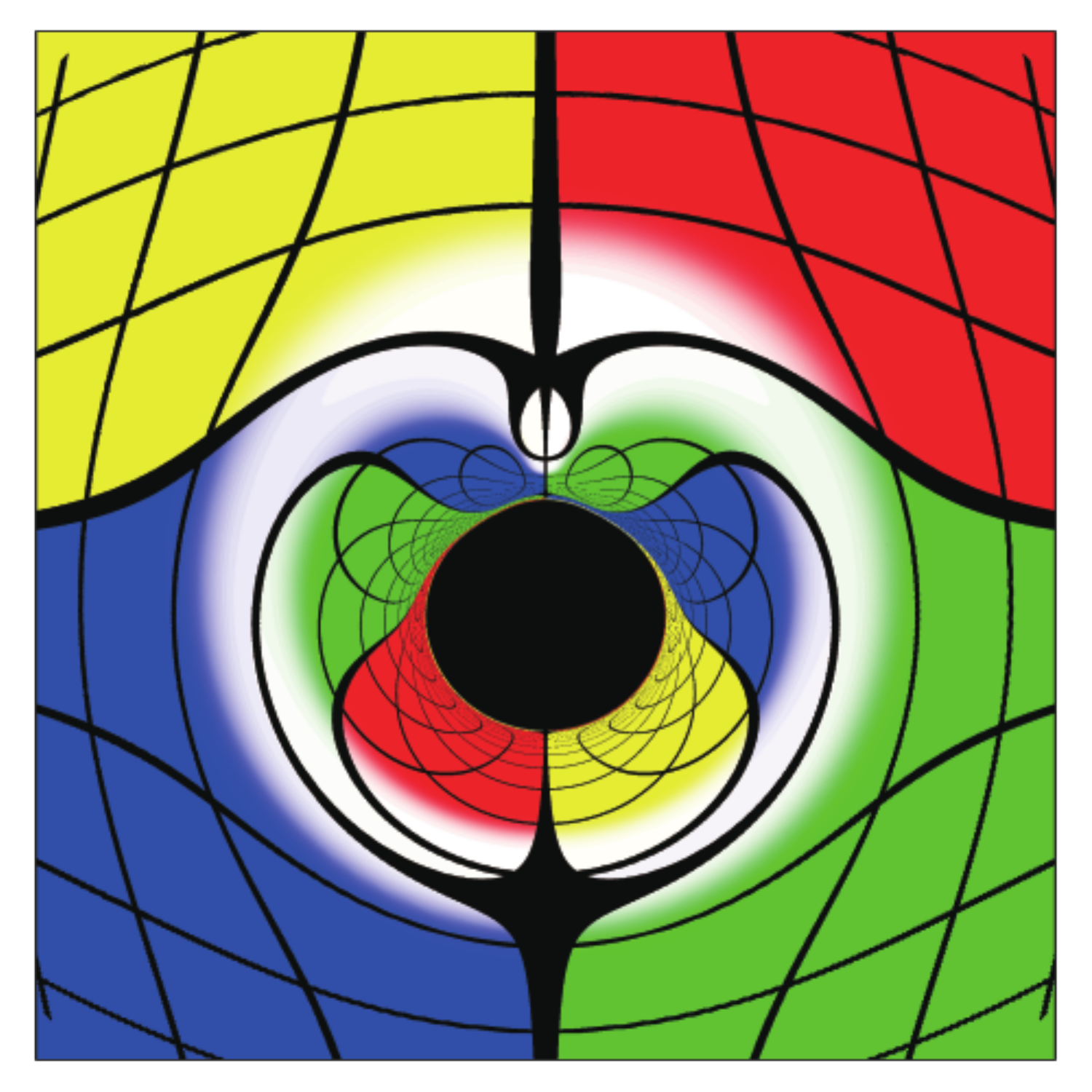}	
	\includegraphics[height=0.19\textwidth,width=0.19\textwidth,angle=90]{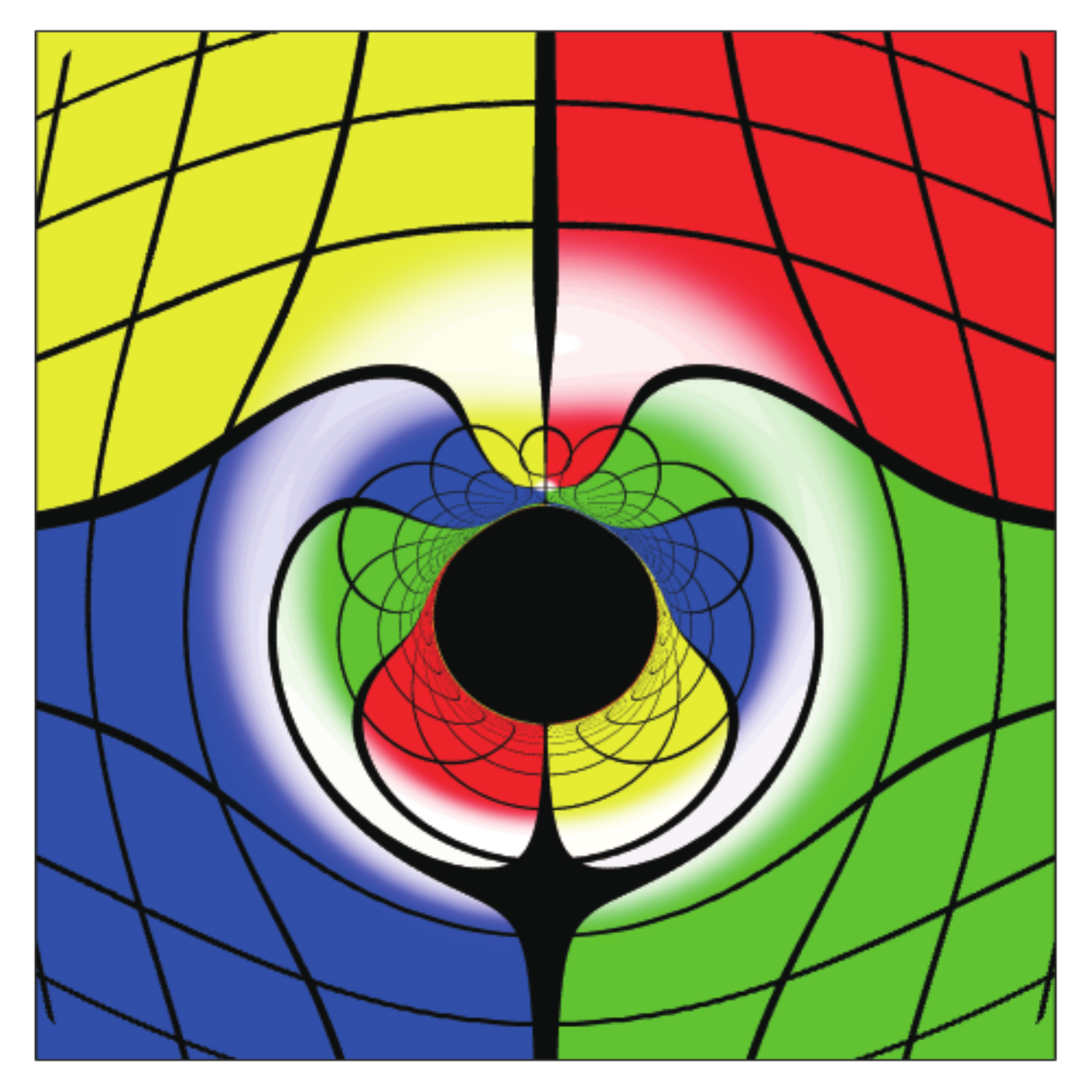}		\includegraphics[height=0.19\textwidth,width=0.19\textwidth,angle=90]{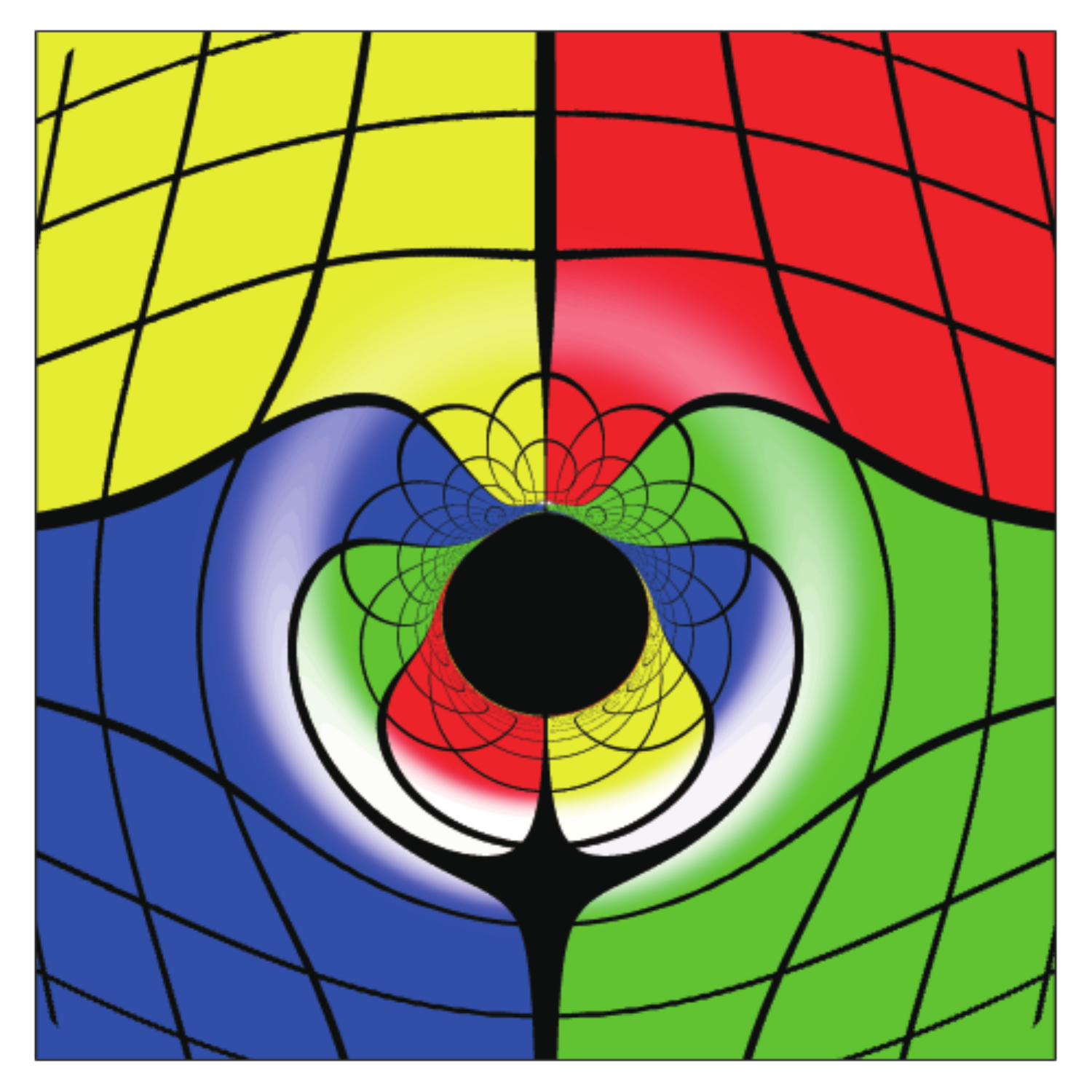}
	\caption{Images of non-rotating ($a=0$, top row) and rotating Kerr ($a=0.998$, bottom row) black holes in a plasma of power-law model. Here, we set $p_0=1$ and choose $h=2$ in Eq.(\ref{Eq: plasma md}). From left to right, $k_0=0,\;3.6^2,\;3.8^2,\;4.0^2,\;4.2^2$.}
	\label{Fig: image in a plasma}
\end{figure}

Taking the power-law model of plasma as an example, in figure \ref{Fig: image in a plasma}, we show the influence of the plasma on the images with 
the background sources formed by the gravitational lensing of the non-rotating (top row) and rotating Kerr (bottom row) black holes. The most left column correspond to a black hole in  vacuum ($k_0=0$).

From the top row, we see that when $k_0$ increases, the Einstein ring of the non-rotating black hole gradually gets close to the shadow, without breaking its symmetry. Inside the Einstein ring are the inverted images of the regions around the white spot, they are squeezed 
in a region near the shadow and are highly distorted, when the Einstein ring moves close to the shadow. 

The images in the bottom row demonstrate the lensing effects for a rotating Kerr black hole. Different from the non-rotating spacetime, the spin of black hole causes frame dragging. This effect is most obvious inside the Einstein ring as shown in the plots. More interestingly, we find that the existence of a plasma breaks the spherical symmetry of the Einstein ring. Furthermore, from the most right image in the bottom row, we see that when the value of parameter $k_0$ is great enough, the image of the white spot would not form a ring.

\begin{figure}
	\centering
	\includegraphics[width=0.4\textwidth,angle=90]{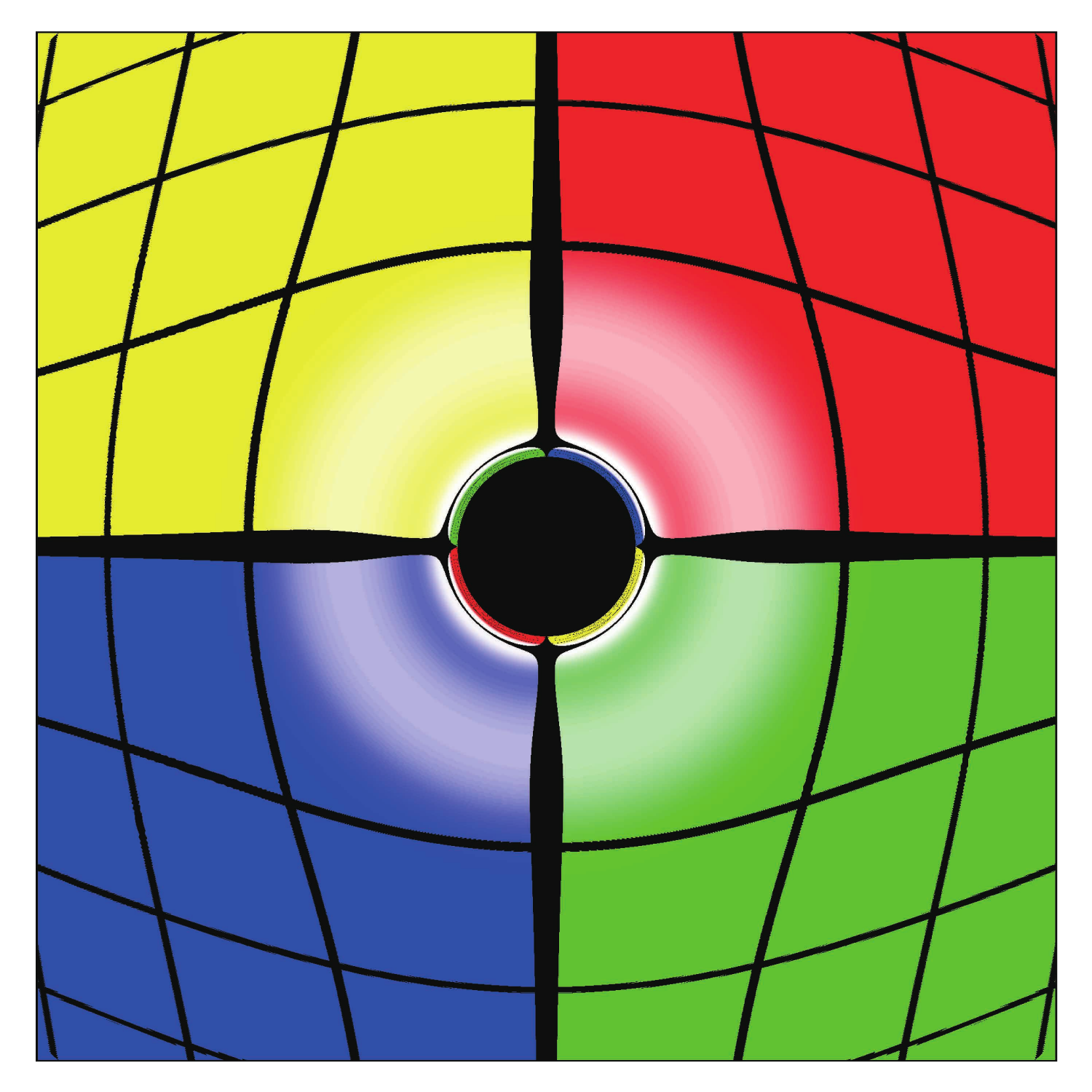}	
	\caption{Multi-ring structure of the image in non-rotating spacetime. Here, we set $p_0=1$ and choose $h=2,\;k_0=4.63^2$ in Eq.(\ref{Eq: plasma md}).}
	\label{Fig: mult_ring}
\end{figure}

As an interesting result,   we remark that a multi-ring structure can be formed in the image of the black hole with the presence of plasma,  due to the refraction of photons with different frequency by the plasma \cite{PhysRevD.87.124009}, see, for a simple example, figure \ref{Fig: mult_ring}.
It is worth noting that, unlike the multi-ring caused by pure gravitational lensing, the outermost ring does not represent the whole image of the white spot and there is no inverted patterns insider the first ring. Some of the photons emitted in the white spot cannot reach the observer's screen.

\section{Conclusions and discussions}
\label{sec:conclusion}
We have analyzed shadows of black holes in the presence of plasma in this paper.  As is pointed out by Perlick and Tsupko\cite{PhysRevD.95.104003}, the Hamilton-Jacobi equations for the motion of photons are not separable and there is no Carter constant, except in some special cases. Therefore, we solved the geodesic equations  numerically to calculate the black hole shadow via a renewed ray-tracing algorithm.

In the presence of a plasma around a black hole, the trajectory of a photon differs from the null geodesic in a way that depends on the photon frequency, resulting in changes of size and shape of the black hole shadow. It is shown that at low frequency or high plasma density, the black hole shadow would disappear in the observer's screen. On the contrary, at high frequency (or in the geometrical-optics limit), the influence of the plasma is negligible and the Kerr shadow is the same as that in vacuum. Moreover, it is also found  that  a  multi-ring structure can be formed in the black hole image due to the presence of a plasma.

As a step towards understanding the influence of  ionized matter on the  various astrophysical compact objects, we here have only considered the photon motion in a stationary, non-magnetized and pressureless plasma in the background of a rotating Kerr black hole. A natural direction for future work would involve extending these results to consider the more realistic but complicated plasma models.

With the upcoming results of the Event Horizon Telescope (EHT), we expect the observation will give us valuable clues and surprise on spacetime and the matter distribution in the neighborhood of black hole. However, it is not advisable to anticipate that EHT is able to constrain the parameters of the plasma density distribution  very well. For a rotating Kerr black hole, although the shadow is flattened in the horizontal direction due to the dragging effect, its vertical diameter can only be shrunk by the effect of plasmas in the context of this work. Furthermore, if the observer is far away from the black hole, its size can be given approximately  by Synge's formula \cite{Grenzebach2015}. For example, let us apply the power-law model of plasma to the Sgr A*, a black hole candidate in our galaxy, whose mass $M=4.3\times 10^6M_{\odot}$ and distance from us $r_{o}=8.3\mathrm{kpc}$. Meanwhile, if we assume the luminosity of the Galactic center $L$ and the accretion efficiency coefficient  are $10^6 L_{\odot}$ and $10^{-4}$, respectively \cite{Narayan1995}, the density parameter $k_0$ in the mm wavelength can be estimated to be $1.2 \times 10^{-4} $ for $h=1.5$. \cite{Perlick:2015vta} Hence, the relative correction to the shadow's vertical diameter due to the plasma is $\sim 4\times 10^{-6}$. Obviously, this is beyond ability of EHT.

\section*{Acknowledgments}
	This work is supported in part by Science and Technology Commission of Shanghai Municipality under Grant No. 12ZR1421700 and Shanghai Normal University.

\bibliographystyle{plain}
\bibliography{Ref}
\end{document}